\documentclass[fleqn,usenatbib]{mnras}

\usepackage{amsmath}

\usepackage[T1]{fontenc}

\usepackage{xcolor}
\usepackage{soul}

\usepackage{comment}
\usepackage{cleveref}
\DeclareRobustCommand{\VAN}[3]{#2}
\let\VANthebibliography\thebibliography
\def\thebibliography{\DeclareRobustCommand{\VAN}[3]{##3}\VANthebibliography}

\usepackage{graphicx}	
\usepackage{amsmath}	
\usepackage{nccmath}
\usepackage{mathtools}
\usepackage{natbib}

\usepackage{booktabs,chemformula}
\usepackage{caption}
\usepackage{array,booktabs,longtable,tabularx}
\usepackage{tabularray}
\usepackage{ltablex}
\usepackage{siunitx}
\usepackage{caption}
\usepackage{fancyhdr}
\usepackage{tocloft}
\usepackage{verbatim}
\usepackage{multicol}
\usepackage{amsmath}
\usepackage{amssymb}
\usepackage{engtlc}
\usepackage{soulutf8}
\usepackage{multirow}
\usepackage{mathtools}
\usepackage{cancel}
\usepackage{subfig}
\usepackage{float}
\usepackage{empheq}
\usepackage[autostyle,italian=guillemets]{csquotes}

\usepackage[normalem]{ulem}
\defcitealias{LCMenegazzi}{Paper~I}

\newcommand{\secref}[1]{\S\ref{#1}}

\usepackage{newtxtext,newtxmath}

\title[]{Variety of disc wind-driven explosions in massive rotating stars. II. Dependence on the progenitor}

\author[Crosato Menegazzi, L, et al.]{
Ludovica Crosato Menegazzi,$^{1}$
Sho Fujibayashi,$^{2,3,1}$
Masaru Shibata,$^{1,4}$
Aurore Betranhandy,$^{1}$
\newauthor
Koh Takahashi$^{5,1}$
\\
$^{1}$Max Planck Institute for Gravitational Physics (Albert Einstein Institute), Am M{\"u}hlenberg 1, Potsdam 14476, Germany\\
$^{2}$Frontier Research Institute for Interdisciplinary Sciences, Tohoku University, Sendai 980-8578, Japan\\
$^{3}$Astronomical Institute, Graduate School of Science, Tohoku University, Sendai 980-8578, Japan\\
$^{4}$Center for Gravitational Physics and Quantum Information, Yukawa Institute for Theoretical Physics, Kyoto University, Kyoto, 606-8502, Japan\\
$^{5}$National Astronomical Observatory of Japan, National Institutes for Natural Science, 2-21-1 Osawa, Mitaka, Tokyo 181-8588, Japan\\
}

\date{Accepted XXX. Received YYY; in original form ZZZ}

\pubyear{2023}

\begin{document}
\label{firstpage}
\pagerange{\pageref{firstpage}--\pageref{lastpage}}
\maketitle

\begin{abstract}
We assess the variance of supernova(SN)-like explosions associated with the core collapse of rotating massive stars into a black hole-accretion disc system under changes in the progenitor structure. Our model of the central engine evolves the black hole and the disc through the transfer of matter and angular momentum and includes the contribution of the disc wind. 
We perform two-dimensional, non-relativistic, hydrodynamics simulations using the open-source hydrodynamic code \texttt{Athena++}, for which we develop a method to calculate self-gravity for axially symmetric density distributions. 
For a fixed model of the wind injection, we explore the explosion characteristics for progenitors with zero-age main-sequence masses from 9 to 40 $M_\odot$ and different degrees of rotation. Our outcomes reveal a wide range of explosion energies with $E_\mathrm{expl}$ spanning from $\sim 0.3\times10^{51}$~erg to $ > 8\times 10^{51}$~erg and ejecta mass $M_\mathrm{ej}$ from $\sim 0.6$ to $> 10 M_\odot$.
Our results are in agreement with some range of the observational data of stripped-envelope and high-energy SNe such as broad-lined type Ic SNe, but we measure a stronger correlation between $E_\mathrm{expl}$ and $M_\mathrm{ej}$. We also provide an estimate of the $^{56}$Ni mass produced in our models which goes from $\sim0.04\;M_\odot$ to $\sim 1.3\;M_\odot$. The $^{56}$Ni mass shows a correlation with the mass and the angular velocity of the progenitor: more massive and faster rotating progenitors tend to produce a higher amount of $^{56}$Ni. 
Finally, we present a criterion that allows the selection of a potential collapsar progenitor from the observed explosion energy.

\end{abstract}




\section{Introduction}
Massive stars ($\gtrsim$ 9 $M_\odot$), at the end of their hydrostatic life, are expected to form an iron core and subsequently undergo a gravitational collapse. 
This core collapse marks the start of a complex sequence of events with various outcomes. The post-collapse evolution and final remnant properties depend on factors like the progenitor mass, its angular momentum, and magnetic field \citep[e.g.,][]{Janka2012PTEP.2012aA309,Ugliano2012ApJ.757.69,Woosley2010ApJ.719L.204}. Typically, stars with moderate mass tend to successfully explode through the heating by neutrinos emitted from the proto-neutron star (PNS), determining a classical core-collapse supernova (CCSN) (e.g., \citealt{Janka_2016,Burrows:2020qrp,2021ApJ...915...28B, Vartanyan:2021dmy, 2022MNRAS.517..543W, Mezzacappa_2020, 2022ApJ...924...38K,Bruenn:2022yoo, Rahman:2023atv} on the latest progress), while progenitors with an even higher zero-age main-sequence mass, $M_\mathrm{ZAMS}\gtrsim16M_\odot$ are more prone to fail the explosion (as indicated by \citealt{Woosley_2006}). The massive stars that fail to launch a successful explosion during the PNS phase collapse into a black hole (BH).

In the presence of an appreciable rotation of the progenitor stars, the BH should be subsequently surrounded by an accreting disc (see, e.g., \citealt{Woosley_2006}). It has been shown that in failed SNe the wind created by the viscous heating inside the accretion disc may be a natural source of the SN energy with an explosion energy $E_\mathrm{expl}\gtrsim10^{52}$ erg (\citealt{macfadyen1999collapsars}, \citealt{Popham_Woosley_1999}, \citealt{Kohri_2005}) and it has been found to be rich in $^{56}$Ni ($\ge 0.1\, M_\odot$) (as shown by \citealt{Just2022, Fujibayashi:2023oyt, Dean_2024}).
The activity of the disc surrounding the newly-born BH can then also be a source of relativistic jets that account for gamma-ray bursts (GRBs).
The BH-disc system is thus a promising engine to explain energetic supernovae such as broad-lined type Ic SNe (Type Ic-BL SNe or hypernovae) and their associated GRBs, as shown by observational studies such as \citet{1998_Galama, Stanek_2003, Campana_2006, Xu_2013} (for a review, see also \citealt{Woosley_Bloom_2006, Kumar_Zhang_2015}). This scenario is known as \textit{collapsar scenario}.

Considering the alternative case of a successful explosion, \citet{Obergaulinger_Aloy_2020} found that PNS with a mass ranging from 1.2 to 2.5 $M_\odot$ can successfully launch explosions through either the neutrino-driven mechanism or the magnetohydrodynamics (MHD)-driven mechanism. 
A MHD-driven CCSN could occur when a strong magnetic field is associated with rapid rotation in the stellar core, and is a possible mechanism for the creation of magnetars. The MHD-driven CCSN scenario (also known as \textit{proto-magnetar} scenario) presents a potential explanation for the GRBs and associated Type Ic-BL SNe (see, e.g., \citealt{Usov_1992}, \citealt{Metzger_2011}). In this scenario, rotation leads to global asymmetries of the shock wave, which translates into the formation of highly collimated, mildly relativistic bipolar outflow as shown by \citet{Burrows_2007}, \citet{Mosta_2015}, \citet{Bugli_2020}, \citet{Obergaulinger_Aloy_2020}, \citet{Kuroda_2020} in their MHD simulations.
\citet{Grimmet_HNe_2021} used hydrodynamics simulations based on this scenario to study the production of $^{56}$Ni. In their most energetic models, where they observed an explosion energy $>10^{52}$ erg, a significant amount of ejected $^{56}$Ni was found, i.e., $>0.05$--$0.45M_\odot$. These findings are consistent with values deduced from the light curves of Type Ic-BL SNe, which range from $0.12$--$0.8M_\odot$, with a median at $\sim 0.28M_\odot$, as determined by \citet{Taddia_2019}. Therefore, both the \textit{collapsar} and \textit{MHD-driven CCSN} scenarios are the currently favored scenarios for the formation of GRBs and associated Type Ic-BL SNe. Historically, scenarios based on neutrino pair annihilation have been discussed through the years (see \citealt{Woosley_1993}, \citealt{2004_Piran} for reviews), but they appear to be less efficient.

This paper is the extension of the previous work we presented in \citet{LCMenegazzi} (hereafter mentioned as Paper I). In the first study, we explored the properties of sub-relativistic outflow in the collapsar scenario, with the explosion fueled by a BH-accretion disc system. We performed two-dimensional axisymmetric hydrodynamics simulations for modeling the ejecta produced by the collapse of the massive, rotating star with $M_\mathrm{ZAMS}=20M_\odot$ taken from \citet{Aguilera-Dena_2020}. Then, by varying the parameters of the injected wind, we investigated their effect on the ejecta properties such as mass, velocity, geometry, and $^{56}$Ni production.
For this progenitor, our analysis unveiled a vast range of explosion energies with $E_\mathrm{expl}$ spanning from very low energy $\sim 5\times10^{49}$~erg to Ic-BL SN energy ($\sim 3\times10^{52}$~erg). This distinction depends on whether the ram pressure of the injected matter is stronger than that of the infalling envelope, effectively pushing the stellar envelope outward or not.
Our results in \citetalias{LCMenegazzi} showed that the explosion energies we measured were in good agreement with observational data for stripped-envelope SNe presented by \citet{Taddia2019jan} and \citet{Gomez2022dec} confirming that the disc wind generated from the BH-disc system in a failed SN may naturally be a source of the SN energy as suggested in previous studies by \citet{Woosley_1993}, \citet{macfadyen1999collapsars}, \citet{Popham_Woosley_1999}. 
Because of these results, we decided to further investigate the variety of the explosion properties using the same scenario as in \citetalias{LCMenegazzi}, but expanding the parameter space by varying the mass and the initial rotational velocity of the progenitor. 
In this study, we fix the parameter of the injected wind while varying the progenitor structure (a detailed description of the choice of the parameters is presented in Section~\ref{sec:Method}). We employ a range of different progenitors with the same composition dominated by oxygen outside the iron core (see \citealt{Aguilera-Dena_2020, Woosley_2006}). Then, we sample them with respect to their $M_\mathrm{ZAMS}$ and vary their degree of rotation, while holding the wind parameters the same.

Our hydrodynamics model based on the collapsar scenario is inspired by those used in fully general relativistic hydrodynamics simulations \citep{Fujibayashi:2023oyt}, which we here simplified. Nonetheless, we will show that this simplified method can reproduce the overall feature of the more detailed simulations, and hence, it is useful for the interpretation of the observational data.

This paper is organized as follows. In \secref{sec:Method}, we first briefly summarize our hydrodynamics code that has been outlined in \citetalias{LCMenegazzi} and we also use for this work, and then we describe the physics of the progenitor stars we employ (taken from \citealt{Aguilera-Dena_2020, Woosley_2006}) and the added parameter for the magnitude of the angular velocity. We present our results in \secref{sec:Results}, where we focus especially on the variety of the explosion energy and the $^{56}$Ni production and study their dependence on the variation of the initial parameters and progenitor models. In this section, we also compare our results with observational data. Then, \secref{sec:Discussion} contains a discussion about the implication of our results, also considering those obtained in \citetalias{LCMenegazzi} and their observational counterpart. Finally, we summarize this work in \secref{sec:Conclusion}. The Appendixes provide an insight into the hydrodynamical evolution of some models excluded from the analysis and an additional study of the effect of the wind injection model on the explosion of the $35M_\odot$ progenitor of \citet{Aguilera-Dena_2020}.  Throughout this paper, $G$ denotes the gravitational constant. 

\section{Method and parameters}\label{sec:Method} 

In this work, we numerically explore the collapsar scenario. Specifically, we focus on massive, fast-rotating stars in which the neutrino-driven explosion during the PNS phase does not take place, causing the PNS to collapse into a BH (failed CCSN). We model the explosion of a compact progenitor star post the BH formation. 
We perform two-dimensional (axisymmetric) Newtonian simulations using the open-source multi-dimensional hydrodynamics code \texttt{Athena++} (\citealt{Stone_2020}) to which we added the self-gravity by solving the Poisson's equation under the axial symmetry (the implementation is presented in \citetalias{LCMenegazzi}). Additionally, in our simulations, we model the central engine in a semi-analytical way by evolving the BH and the disc through the transfer of matter and angular momentum. This method follows the prescriptions given by \citet{Kumar_Narayan_method} on which we add the contribution of the disc as described by \citet{Hayakawa_2018}.

The thermodynamical history of the ejecta is obtained using tracer particles in the method provided in~\citetalias{LCMenegazzi}. This method allows us to follow the evolution of the tracer particles backward in time and also to distinguish the fluid elements of the injected matter (coming from the inner boundary) from those originating from the stellar envelope (for a detailed description, see \citetalias{LCMenegazzi}). 
If the maximum temperature of a tracer particle is higher than the critical temperature 5~GK for nuclear statistical equilibrium (e.g., \citealt{Woosley2002RvMP.74.101}), we assume that $^{56}$Ni is synthesized with mass assigned to the tracer particle.
As for the injected matter, since it lacks the thermodynamical history, we estimate the mass of $^{56}$Ni by evaluating the temperature of the disc when the matter is injected. We, then, measure the ratio between the injected matter experiencing temperature higher than 5~GK and the total injected mass and we multiply it to the injected mass to get the amount of $^{56}$Ni produced in this component of the ejecta. An explanation of the procedure used to estimate the temperature of the injected matter is presented in Appendix ~\ref{Appendix:T_inj_matter}.
We do not perform a full nucleosynthesis calculation and use the critical temperature \SI{5}{GK} to approximately estimate the $^{56}$Ni mass produced in the ejecta because of the lack of knowledge about the injected matter and the fact that it dominates over the mass originating from the stellar envelope (see \S~\ref{sec:Results:Ni_production}).

\begin{figure}
	\centering
	\includegraphics [width=0.48\textwidth]{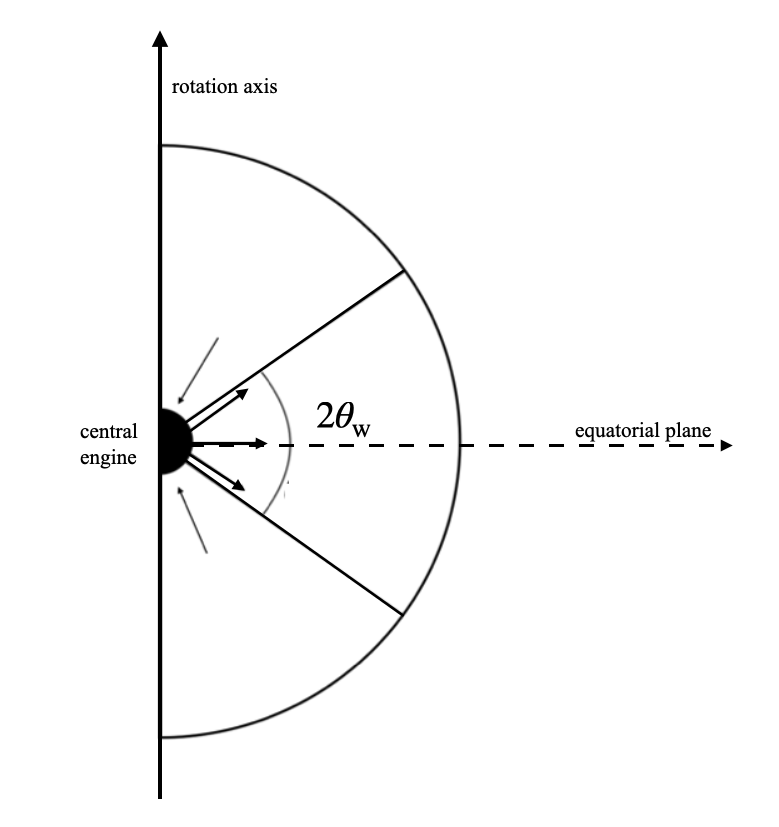}
	\caption{Schematic picture of the explosion in the collapsar scenario. $2\theta_\mathrm{w}$ represents the angle for which we allow the wind outflow. Outside this angle, the matter is only allowed to infall towards the central engine.
 The figure also shows the rotation axis and the equatorial plane.}
    \label{fig:scheme_star}
\end{figure}

\subsection{Computational setup}

In this work, all simulations are performed on an axisymmetric grid using spherical-polar coordinates. The polar angle of our domain spans from 0 to $\pi$ with $128$ grid points uniformly distributed, leading to a zone width of $\Delta\theta=0.0245$ rad. The radial dimension ranges from $10^8$ cm ($r_\mathrm{in}$) to $3.3\times 10^{10}$ cm ($r_\mathrm{out}$) and it is divided into $220$ zones. The grid zone size $\Delta r$ is obtained by increasing the mesh size with a constant factor $\Delta r_i = \alpha \Delta r_{i-1}$ where $\alpha\approx 1.03$ ensures an approximately squared shape for all zones (i.e., $\Delta r_i \approx r_i\Delta \theta$; for more details see \citetalias{LCMenegazzi}).

The inner radius $r_\mathrm{in}$ determines the inner boundary of the computational domain, and it is the same through all simulations. This cut is done to exclude the central engine from the computational domain and consider it embedded in the central part of the star. By this cut, computational costs are significantly saved, as evolving the central engine semi-analytically rather than numerically significantly reduces the simulation time.

Instead of solving the hydrodynamics inside $r_\mathrm{in}$, we evolve the BH and disc assumed to be embedded there. Specifically, their masses ($M_\mathrm{BH}$ and $M_\mathrm{disc}$) and angular momenta ($J_\mathrm{BH}$ and $J_\mathrm{disc}$) are evolved according to the mass and angular momentum fluxes at $r_\mathrm{in}$ (see \citetalias{LCMenegazzi} for details).
The initially enclosed mass and angular momentum inside $r_\mathrm{in}$ are assumed to be those of the initial mass of the BH (see \S~\ref{sec:Inner_Boundary_Cond}). The outer radius, which is located outside the stellar surface of our progenitor models, is also kept fixed for all simulations.

\subsection{The equation of state}
The thermodynamical properties of the star are described by the same equation of state (EOS) as employed in \citetalias{LCMenegazzi}
which includes ions, radiation, electrons, and $e^-$-$e^+$ pair (we also refer to the EOS described in \citealt{Timmes_2000} and \citealt{Takahashi_2016}). In this work, we suppose that oxygen is the only component for the ion (i.e., $Y_{e}=0.5$), resulting in a $^{16}$O mass fraction of 1. This decision was made considering the composition of our progenitor model, which is dominated by oxygen outside the iron core (see \citealt{Aguilera-Dena_2020}).

\subsection{Inner boundary condition and Parameters}\label{sec:Inner_Boundary_Cond}
The model used in this work is the same as in \citetalias{LCMenegazzi}, which is based on the collapsar disc wind scenario and inspired by \citet{Fujibayashi:2023oyt} (see also \citealt{Just2022} and \citealt{Dean_2024}). We set the inner boundary conditions so that the wind is launched after the disc formation toward the non-axis direction (see Sec.2.6 of \citetalias{LCMenegazzi} for a detailed description of the inner boundary conditions).  
In the early stages, before the disc formation, we allow material to flow toward the central engine for $r < r_\mathrm{in}$ (outflow condition). 
After the disc formation ($M_\mathrm{disc}>0$), we set the wind injection from the inner boundary within a half opening angle of $\theta_\mathrm{w} = \pi/4$ directed along the equatorial plane. Outside the injection angle, we inhibit matter flowing from the central engine into the computational domain by setting zero fluxes (reflecting boundary condition) when the radial velocity in the first active cell is positive while letting the mass infall to the central engine if it is negative. The geometry employed in our simulations is illustrated in Fig. \ref{fig:scheme_star}.

We set the wind density $\rho_\mathrm{w}$ using a parabolic density profile described in \citetalias{LCMenegazzi}.
The total specific energy of the disc wind at the inner boundary is assumed to be a fraction of the specific kinetic energy with the disc escape velocity $v_\mathrm{esc}$ as:
\begin{align}
    \frac{1}{2}v_{\mathrm{w}}^2 +f_\mathrm{therm} \frac{1}{2}v_{\mathrm{w}}^2+\Phi = \xi^2 \frac{1}{2}v_\mathrm{esc}^2 \;\label{e_inj_eqs},
\end{align}
where $v_\mathrm{esc}:=\sqrt{2GM_\mathrm{BH}/r_\mathrm{disc}}$ with $r_\mathrm{disc}:=j_\mathrm{disc}^2/GM_\mathrm{BH}$ and $j_\mathrm{disc} := J_\mathrm{disc}/M_\mathrm{disc}$. 
In equation~\eqref{e_inj_eqs}, $\Phi$ is the gravitational potential which satisfies the Poisson's equation:
\begin{align}\label{Poisson_eq}
    \Delta \Phi=4\pi G\rho \; ,
\end{align}
where $\rho$ is the mass density of the fluid, and the specific internal energy of the wind, $e_{\mathrm{int,w}}/\rho_\mathrm{w}=(1/2)f_\mathrm{therm} v_{\mathrm{w}}^2$, is defined as a fraction of the wind kinetic energy through the free parameter $f_\mathrm{therm}$. The fudge factor $\xi$ denotes the uncertainties due to incomplete knowledge of the disc structure \citep{Hayakawa_2018}.
The outflow pressure is calculated from the EOS with the density ($\rho_\mathrm{w}$) and internal energy ($e_{\mathrm{int,w}}$) as input parameters. 

Equation~\eqref{e_inj_eqs} indicates that if the total specific energy $(1/2)v^2+e_\mathrm{int}/\rho+\Phi$ is conserved, the asymptotic velocity of the injected matter is $\xi v_\mathrm{esc}$.

Part of the injected matter could fall back to the center when it has a ram pressure smaller than that of the infalling envelope. If this happens and the injected matter that is pushed back has $j>j_\mathrm{ISCO}$ where $j_\mathrm{ISCO}$ is the specific angular momentum at the innermost stable circular orbit (ISCO) \citep{Bardeen_ISCO}, it would become part of $M_\mathrm{disc}$ again and hence re-injected into the computational domain. To avoid the recycling of the injected matter, as we did in \citetalias{LCMenegazzi}, we do not allow the injected matter to fall back to the disc, but only to the BH by setting the angular momentum of the injected matter to zero.

In this work, we fix the parameters of the wind following the results obtained in our previous work. The wind parameters are the wind timescale $t_\mathrm{w}$, the accretion time scale $t_\mathrm{acc}$, the ratio between the radial velocity of the outflow and the escape velocity $\xi$, and $f_{\mathrm{therm}}$ (see equation~\eqref{e_inj_eqs}). This regulates the rate at which material is accreted onto the black hole from the disc, and hence aids in monitoring the central engine's dynamics (see \citealt{Kumar_Narayan_method} for more details). 

Our aim of the present study is to reproduce a highly energetic explosion with a large production of $^{56}$Ni, comparable to observed high-energy SNe. Therefore, referring to Figure~7 of \citetalias{LCMenegazzi}, all simulations in this work are performed setting a wind time scale $t_\mathrm{w} = 3.16$ s, the accretion timescale $t_\mathrm{acc}=10$~s, the factor $\xi^2 = 0.1$ and $f_\mathrm{therm} = 0.1$ (but see Appendix~\ref{Appendix:model_35_M} for a complementary study).  These parameters characterized the model M20\_3.16\_3.16\_0.1\_0.10 of \citetalias{LCMenegazzi}, which undergoes an energetic explosion with $E_\mathrm{expl} = 3.0\times \SI{e51}{erg}$ and has the ejecta mass of $M_\mathrm{ej}= 3.4\,M_\odot$, which is also in good agreement with the results obtained by \citet{Fujibayashi:2022xsm} for the same progenitor\footnote{\citet{Fujibayashi:2022xsm} measured the ejecta mass of $M_\mathrm{ej}= 2.2\, M_\odot$ and an explosion energy of $E_\mathrm{expl}=2.2\times10^{51}$ erg at the end of their simulation. These values are lower than ours, but they should be considered as the lower limits since they were still growing at the end of their simulation; in a longer-term simulation, these values can be larger.}.

\subsection{Diagnostics}\label{sec:Diagnostic}
In this subsection, we outline the approach used to calculate the properties of the ejecta and injected matter. We define ejecta mass $M_\mathrm{ej}$ as the sum of unbound matter mass.
The explosion energy $E_{\mathrm{expl}}$ is the energy carried by the unbound matter. The unbound matter is evaluated through the Bernoulli criterion, which takes into account the thermal effect on the matter and the effects of the gravitational potential, and is defined as:
\begin{align}
    B := \frac{e_\mathrm{t}+P}{\rho}+\Phi>0, \label{Bernoulli_criterion}
\end{align}
where $e_\mathrm{t} = e_\mathrm{int}+e_\mathrm{kin}$ is the sum of the internal ($e_\mathrm{int}$) and kinetic ($e_\mathrm{kin}=\rho v^2/2$) energy densities, and $P$ is the pressure of the fluid, respectively.

Using the Bernoulli criterion, we track the evolution of the ejecta mass and energy at every time step by integrating the equations:
\begin{align}
M_\mathrm{ej} = r_\mathrm{out}^2 \int_0^t \int_{B>0,v_r>0} \rho v_r dS dt + \int_{B>0,v_r>0} \rho d^3x ,
\end{align}
\begin{align}
E_\mathrm{expl} & =  r_\mathrm{out}^2 \int_0^t \int_{B>0,v_r>0} \rho B v_r dS dt +\int_{B>0,v_r>0} (e_\mathrm{t} +\rho\Phi) d^3x , 
\end{align}
where, $dS$ denotes the surface integral element.
The injected mass $M_\mathrm{inj}$ represents the matter coming from the central engine with a positive mass flux at the inner boundary $r_\mathrm{in}$. It is defined as:
\begin{align} \label{M_injected_def}
    M_\mathrm{inj} = r_\mathrm{in}^2 \int_0^t \int_{B>0,v_r>0} \rho v_r dS dt.
\end{align}

We consider the injected energy $E_\mathrm{inj}$ as the energy carried by $M_\mathrm{inj}$ with positive binding energy. 
We, then, compute the injected energy $E_\mathrm{inj}$ applying the Bernoulli criterion as follows:
\begin{align}
E_\mathrm{inj}  =  r_\mathrm{in}^2 \int_0^t \int_{B>0,v_r>0} \rho B v_r dS dt.
\end{align}

The Bernoulli criterion allows us also to evaluate the energy of the bounded matter (i.e., the binding energy) at the beginning of the injection which we define as:
\begin{align}
E_\mathrm{bind,inj} & = \int_{B<0,v_r<0} (e_\mathrm{t} +\rho\Phi) d^3x. \label{E_bind}
\end{align}

\subsection{Progenitor Parameters}\label{sec:Progenitor_Parameters}

Progenitors of long GRBs have been suggested to be rapidly rotating, rotationally mixed massive stars.  
Therefore, based on our aforementioned aim of investigating the effect of the progenitor structure on the explosion properties, in this work, we fix the parameters of the wind injection (explained in Sec.\ref{sec:Inner_Boundary_Cond}) while changing the progenitor models. We, then, select some massive, rapidly rotating, rotationally mixed stars from the stellar evolution models of \citet{Aguilera-Dena_2020} (throughout this paper we will refer to the $M_\mathrm{ZAMS}$ of the model also as $M_\mathrm{prog}$). In particular, we choose nine stars with $M_\mathrm{ZAMS}=$ 9, 15, 17, 20\footnote{This is the same progenitor used in \citetalias{LCMenegazzi}.}, 22, 25, 30, 35 and 40 $M_\odot$. 
For each of them we, then, consider five different degrees of rotation, the original rotational profile of the progenitor $\Omega_0$ (as given by \citealt{Aguilera-Dena_2020}) and four more for which the angular velocity is modified from the original one $\Omega_0$. These are obtained multiplying $\Omega_0$ by $n_\Omega = 0.5,\,0.6,\, 0.8$, and $1.2$ following \cite{Fujibayashi:2023oyt}.

In order to further investigate the dependence of the ejecta mass, the explosion energy, and the $^{56}$Ni production on the progenitor structure, we also perform additional simulations using models with different characteristics. Specifically, we employ the progenitor models \texttt{16TI} and \texttt{35OC} from \citet{Woosley_2006} with no modification of the angular velocity. These stars have a larger angular momentum in an inner region than those of \citet{Aguilera-Dena_2020} at the onset of the stellar collapse.

Considering the progenitor stars of \cite{Aguilera-Dena_2020}, in our model selection, we include both progenitors that are expected to undergo a successful explosion during the PNS phase and progenitors that are expected to fail the explosion and lead to the BH formation, according to the core-compactness criterion (\citealt{Ertl_2016} and \citealt{Muller_2016}: The progenitors' core-compactness parameter is presented in Figure~4 in \citealt{Aguilera-Dena_2020}). This choice is made in order to cover a wide range of compactness of the \textit{entire} star, which is likely to be relevant to the mass accretion rate after the disc formation (see \citealt{Fujibayashi:2023oyt}). For instance, higher compactness of the entire star leads to a higher mass accretion rate at a later stage of the collapse (i.e., after the disc formation), and \citet{Fujibayashi:2023oyt} showed that a higher mass-infall rate, typically from the carbon-oxygen layer of the star, amplifies the viscous and shock heating rates within the inner region of the disc, determining a large explosion energy. Therefore, studying different values of the entire star compactness allows us to investigate a large variety of mass accretion rates after the disc formation, which, in our scenario, may affect the outflow energy. 

\begin{figure}
	\centering
	\includegraphics [width=0.48\textwidth]{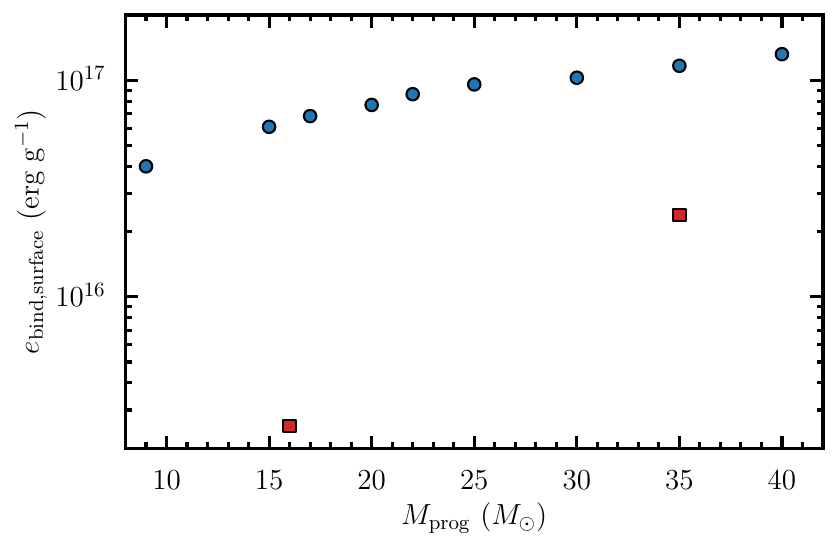}   
	\caption{The specific gravitational binding energy at the surface of the progenitor star $e_\mathrm{bind,surface}$ as a function of the progenitor mass $M_\mathrm{prog}$. $e_\mathrm{bind,surface}$ gives an estimate of the compactness of the entire star. The blue circles show $e_\mathrm{bind,surface}$ for the models of \citet{Aguilera-Dena_2020}, while the red squares indicate the results obtained using \texttt{16TI} and \texttt{35OC} (\citet{Woosley_2006}.} 
    \label{fig:star_compactness}
\end{figure}

A measure of the compactness of the entire progenitor star can be given by estimating the specific gravitational binding energy at the surface of the star defined by:
\begin{equation}\label{eq:Method:E_bind_prog}
    e_\mathrm{bind,surface}=\frac{GM_*}{R_*} \; ,
\end{equation}
where $M_*$ and $R_*$ are the mass and the radius of the star, respectively. 
In Figure~\ref{fig:star_compactness} we show the value of $e_\mathrm{bind,surface}$ as a function of the progenitor mass for the models of \citet{Aguilera-Dena_2020} (blue circles) and the \texttt{16TI} and \texttt{35OC} progenitors from \citet{Woosley_2006} (red squares). The figure shows that more massive stars have higher specific gravitational binding energy considering the models of \citet{Aguilera-Dena_2020} because $R_*$ depends only weakly on $M_*$ for their models. By contrast, for the models of \citet{Woosley_2006}, $R_*$ is much larger than that for the models of \citet{Aguilera-Dena_2020}, and thus, the specific gravitational binding energy is much smaller. This suggests that the models of \citet{Woosley_2006} have the possibility of more energetic explosion.

\begin{figure*}
\centering
    \includegraphics [width=0.48\textwidth]{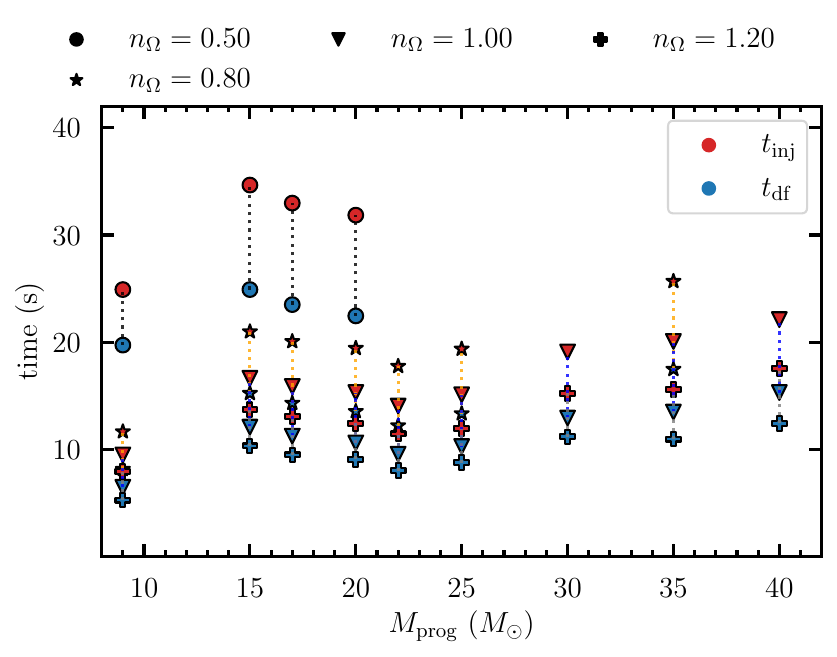} 
    \includegraphics [width=0.48\textwidth]{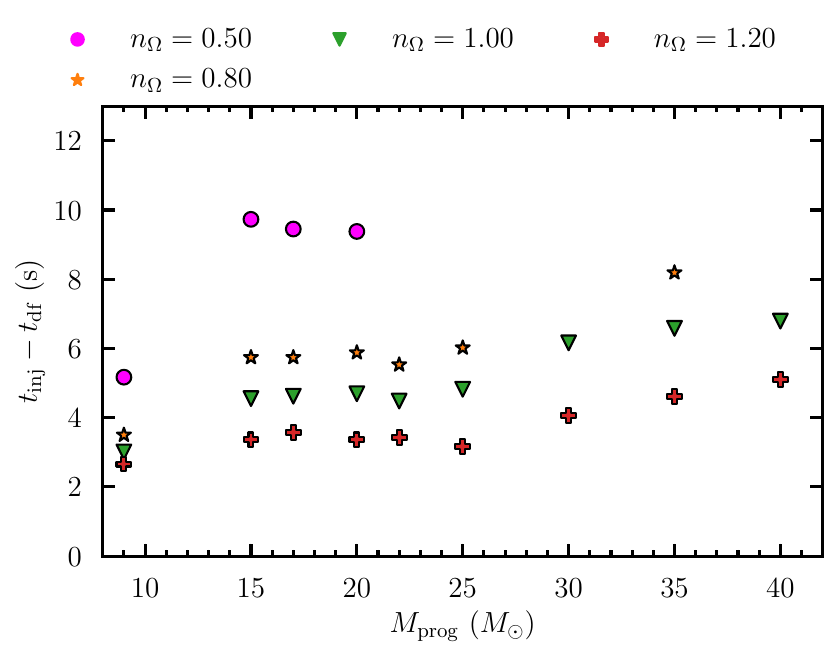}   
    \caption{Left panel: disc formation time, $t_\mathrm{df}$ (blue dots), and injection time (red dots), $t_\mathrm{inj}$, as functions of the progenitor mass for a sample of models taken from \citet{Aguilera-Dena_2020} with four degrees of rotation: $n_\Omega = 0.5$, $0.8$, $1$ and $1.2$. Right panel: time difference between the disc formation and the beginning of the injection, $t_\mathrm{inj} -t_\mathrm{df}$ for the same four rotational levels. In this panel the color and the shape distinguish the magnitude of the angular velocity. 
    }
    \label{fig:t_inj_t_df_Omega}
\end{figure*}

\section{Results}\label{sec:Results}
In this section, we will first analyze the mechanism that drives the explosion in our models and will investigate the dependence of the explosions on the progenitor mass and angular velocity (\S~\ref{sec:Results:dynamics}). We also compare those to some observational data (\S~\ref{sec:Results:Comparison_obs}). We then present a way to predict the explosion energy from our progenitor models (\S~\ref{sec:Results:explodability_prediction}). Finally, we will show the effect of the progenitor mass $M_\mathrm{prog}$ and the magnitude of the angular velocity $n_\Omega$ on the final $^{56}$Ni production (\S~\ref{sec:Results:Ni_production}).
The models studied in this work, with the most important properties of their ejecta and the results of the $^{56}$Ni production, are summarised in Table~\ref{tab:models}. 

In the final analysis of the results, some models were excluded due to their physical inconsistency. The inconsistency arises from the fact that in these models, the accretion disc grows in the computational domain of $r_\mathrm{in} \leq r \leq r_\mathrm{out}$. In our numerical modeling there are no processes that launch the wind from the disc in the computational domain. Consequently, the injected mass and energy are not correctly assessed for such models. On the other hand, for the other models, the mass and energy injection sets in before the disc is formed in the computational domain, and thus the result is considered to be physically consistent. 
Models showing the inconsistent behavior are therefore excluded from the general analysis of the results presented here but are described in Appendix~\ref{Appendix:exceptional_models} and listed in Table~\ref{tab:models_exceptions}. 

\begin{table*}
\centering
\caption{Model description and key results. The model's name contains information about the progenitor mass and the magnitude of the angular velocity: the first number corresponds to $M_\mathrm{prog}$ and the second indicates the factor by which the original degree of rotation has been multiplied.  From left to right, the columns list the cumulative injected energy, ejecta mass, explosion energy, average ejecta velocity, the number of tracer particles located within the ejecta, the mass of ejecta component originating from the injected matter, the ratio between the injected matter that is estimated to experience temperature higher than 5 GK and the total injected mass,
the mass of ejecta component that is originated from the computational domain and experiences temperature higher than 5\,GK, along with the number of tracer particles in parenthesis, the total mass of the ejecta which experiences temperature higher than 5\,GK. 
}
\begin{tabular}{c|ccccccccc}
\hline\hline
model & $E_\mathrm{inj}$ & $M_\mathrm{ej}$ & $E_\mathrm{expl}$ & $v_\mathrm{ej}$ & $N_\mathrm{p}$ & $M_\mathrm{ej}^\mathrm{inj}$ & $M_\mathrm{>5GK}^\mathrm{inj}/M^\mathrm{inj}$ & $M^\mathrm{stellar}_\mathrm{ej,>5GK}$ ($N^\mathrm{stellar}_\mathrm{ej,>5GK}$) &  $M_\mathrm{ej,>5GK}$ \\
& ($10^{51}$\,erg) & ($M_\odot$) & ($10^{51}$\,erg) & ($10^3$\,km/s) &      & ($M_\odot$)      &       &  ($M_\odot$) &($M_\odot$)   \\ 
\hline
AD009x0.5 &  0.59     &  0.56   &    0.26  &  6.67      &   38681 & 0.19 & 0.39  &  0.00 (     0)  &  0.07\\
AD015x0.5 &  2.51     &  1.49   &    0.93  &  7.79      &   4269 & 0.46 & 0.44  &  0.01 (   160)   &  0.21\\
AD017x0.5 &  2.81     &  1.52  &    0.91  &  7.61      &   43049 & 0.53 &  0.44 &  0.01 (   111)   &  0.24\\
AD020x0.5 &  3.37     &  1.36   &    0.86  &  7.82      &   37600 & 0.51 &  0.46 &  0.01 (   145)  &  0.24\\

\hline
AD009x0.6 &  0.93     &  0.78  &    0.35  &  6.72      &   52445 & 0.10 &  0.43 &  0.00 (     1)  &   0.04\\
AD015x0.6 &  3.15     &  1.88   &    1.34  &  8.32      &   41193 & 0.64 & 0.48  &  0.02 (   187)  &   0.33\\
AD017x0.6 &  3.63     &  1.94   &    1.43  &  8.46      &   43309 & 0.65 &  0.48 &  0.017 (   211)     &  0.33 \\
AD020x0.6 &  4.67     &  1.77   &    1.35  &  8.61      &   39282 & 0.50 &  0.50 &  0.01 (   165)  &  0.26\\
AD022x0.6 &  5.80     &  2.00   &    1.95  &  9.74      &   61595 & 0.75 &   0.52 &  0.01 (   123)  &   0.40\\
AD025x0.6 &  6.22     &  1.85   &    1.60  &  9.17      &   43669 & 0.54 &  0.54 &  0.02 (   249)  &   0.31\\

\hline
AD009x0.8 &  1.58     &  1.12   &    0.78  &  8.24      &   59195 & 0.20 &  0.54 &  <0.01 (    12)    &   0.12\\
AD015x0.8 &  3.82     &  2.43   &    1.61  &  8.01      &   39308 & 0.78 &  0.44 &  0.02 (   208)  &   0.36\\
AD017x0.8 &  4.71     &  2.50   &    2.04  &  8.91      &   45816 & 0.83 &  0.52 &  0.04 (   353)  &   0.47\\
AD020x0.8 &  6.45     &  2.28   &    1.98  &  9.20      &   42409 & 0.54 & 0.50 &  0.03 (   277)  &   0.30\\
AD022x0.8 &  7.91     &  3.03   &    3.36  &  10.39     &   43355 & 1.20 &  0.55 &  0.02 (   136)  &  0.68\\
AD025x0.8 &  9.77     &  2.82   &    3.13  &  10.63     &   89755 & 0.88 &  0.64 &  0.02 (   138)  &   0.58\\
AD035x0.8 &  11.87    &  2.80   &    2.86  &  9.97      &   64595 & 1.11 &  0.55 &  <0.01 (    24) &   0.62\\
\hline
AD009x1.0 & 2.03      &  1.39   &    1.11  &  8.82      &   45629 & 0.32 &  0.64 &  <0.01 (     8)    &   0.21\\
AD015x1.0 & 4.21      &  2.37   &    1.37  &  7.50      &   41690 & 0.53 & 0.45 &  0.02 (   170)  &   0.26\\
AD017x1.0 & 5.21      &  3.07   &    2.21  &  8.36      &   40497 & 1.09 &   0.48 &  0.05 (   344)  &  0.57\\
AD020x1.0 & 7.10      &  3.33   &    3.04  &  9.42      &   45856 & 1.23 &   0.52 &  0.06 (   380)  &   0.70\\
AD022x1.0 & 9.05      &  3.61   &    3.97  &  10.28     &   49035 & 1.44 &  0.56 &  0.04 (   278)  &  0.85\\
AD025x1.0 & 11.50     &  3.94   &    4.75  &  10.82     &   49748 & 1.52 & 0.59 &  0.05 (   257)  &   0.95\\
AD030x1.0 & 13.46     &  4.40   &    5.16  &  10.68     &   55761 & 1.88 &  0.54 &  0.02 (    57)  &   1.04\\
AD035x1.0 & 16.20     &  4.30  &    5.15  &  10.80     &   53789 & 1.70 &  0.56 &  <0.01 (    28) &  0.96\\
AD040x1.0 & 17.55     &  3.73   &    3.38  &  9.37      &   38332 & 1.36 &   0.57 &  0.10 (   663)  &  0.88\\
\hline
AD009x1.2 & 2.38      &  1.56   &    1.35  &  9.16      &   45537 & 0.39 & 0.65 &  <0.01 (    14)     &   0.26\\
AD015x1.2 & 4.19     &  2.98   &    1.74  &  7.52      &   62670 & 0.82 &   0.45 &  0.01 (    83)   &  0.38\\
AD017x1.2 & 5.55      &  3.28   &    2.25  &  8.17      &   39335 & 1.07 &  0.49 &  0.04 (   294)  &   0.56\\
AD020x1.2 & 7.83      &  3.75   &    3.46  &  9.45      &   61895 & 1.24 &  0.53 &  0.06 (   378)  &   0.72\\
AD022x1.2 & 9.77      &  4.07   &    4.16  &  9.96      &   72141 & 1.49 &  0.56 &  0.08 (   381)  &   0.91\\
AD025x1.2 & 13.38     &  4.74   &    5.70  &  10.81     &   65685 & 1.72 &   0.58 &  0.10 (   445)  &  1.10\\
AD030x1.2 & 15.00     &  5.27   &    6.19 &  10.84     &    49952 & 1.98 &  0.56 & 0.093 (396)  &   1.20 \\
AD035x1.2 & 19.73     &  5.58   &    7.98  &  11.79     &   46960 & 2.21 &   0.55 &  0.04 (   172)  &  1.26\\
AD040x1.2 & 22.86     &  5.64   &    8.15  &  11.82    &   58541 & 2.28 &   0.55 &  0.017 (    34)  &  1.27\\
\hline
16TI      & 3.28     &  5.32   &    1.20  &   4.76  & 87957 & 0.60  & 0.41 & 0.015 (    1261) &  0.26\\
35OC     & 14.42     &  10.67  &     5.18  &  6.99   & 145059 & 1.88 & 0.47 & 0.25 (    11268)& 1.14 \\
\hline
\end{tabular}
\label{tab:models}
\end{table*}

\subsection{Parameter dependence of the dynamics}\label{sec:Results:dynamics} 
To investigate the dependence of the explosion properties on the progenitor models we fixed the wind parameter to be the same for all simulations. As explained in \S~\ref{sec:Inner_Boundary_Cond}, we employed the wind parameter of the model M20\_3.16\_3.16\_0.1\_0.10 from \citetalias{LCMenegazzi} (i.e., $t_\mathrm{w} = 3.16$ s, $t_\mathrm{acc}/t_\mathrm{w} = 3.16$, $\xi^2 = 0.1$ and $f_\mathrm{therm} = 0.1$) because it undergoes an energetic explosion with explosion energy and $^{56}$Ni mass comparable to those measured in energetic SNe.

In the left panel of Figure~\ref{fig:t_inj_t_df_Omega}, we compare the disc formation time $t_\mathrm{df}$, at which $M_\mathrm{disc}$ becomes non-zero, with the injection time $t_\mathrm{inj}$ (the time at which the mass and energy fluxes at the inner boundary become positive and the matter unbound) for a sample of progenitors (among those taken from \citet{Aguilera-Dena_2020}) with four different rotational levels.

Both panels of Figure~\ref{fig:t_inj_t_df_Omega} show a correlation between the disc formation time and the progenitor structure, specifically, the magnitude of the angular velocity: $t_\mathrm{df}$ is longer for progenitors with lower values of $n_\Omega$. It is possible to explain this behavior considering that in slower rotating models the matter that has a sufficient angular momentum to form the disc is located at larger radii, thus taking longer time to fall into the central region and form the disc.

To prove the above speculation, we evaluate the free-fall time of the mass shell that has a specific angular momentum sufficiently large to form the disc, which may give an approximate estimation of $t_\mathrm{df}$.
We here assume that during the formation and the growth of a BH, the specific angular momentum is conserved. Under this assumption and considering that the region with the enclosed mass $M_\mathrm{encl}$ collapses to the BH without forming a disc, it is possible to estimate the mass and angular momentum of the formed BH for a given profile of the specific angular momentum as a function of the enclosed mass, $j(M_\mathrm{encl})$ \citep{Shibata_2002, Shibata_2003}, defined as
\begin{align}
j = \frac{1}{4\pi r^2}\int_0^{2\pi}\int_0^{\pi}\Omega(r)r^4\sin^3\theta d\theta d\phi = \frac{2}{3}r^2\Omega(r) \label{j_average_t_ff}.
\end{align}
Here, $\Omega(r)$ is the angular velocity profile as a function of the spherical radius only, which is assumed in stellar evolution calculations~\citep{Zahn1992}. Both $j$ and $M_\mathrm{encl}$ are functions of $r$.

\begin{figure}
\centering
\includegraphics [width=0.48\textwidth]{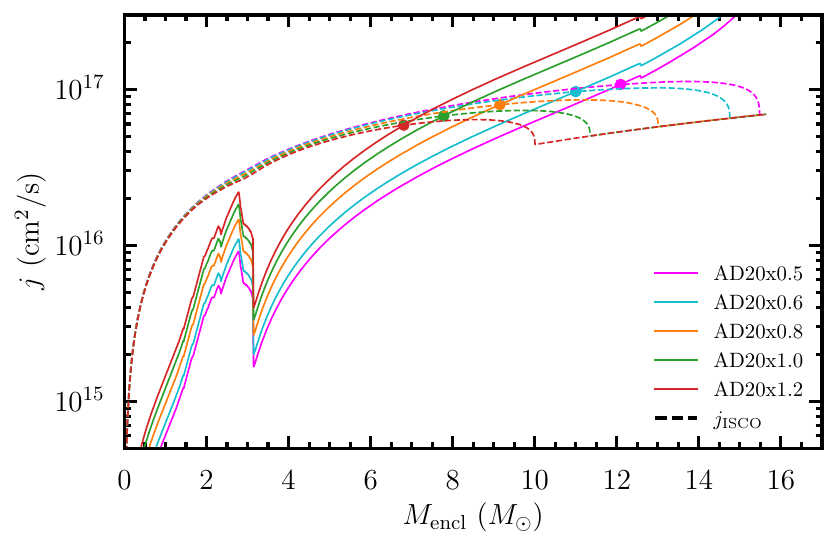}
\caption{Specific angular momentum, $j$, as a function of the enclosed mass, $M_\mathrm{encl}$, for the model of $M_\mathrm{prog} = 20\, M_\odot$ with different degrees of rotation: $n_\Omega = 0.5, 0.6, 0.8, 1.0$ and $1.2$. The magnitude of the angular velocity is distinguished by the color. We also plot $j_\mathrm{ISCO}$ for a given BH of mass $M_\mathrm{encl}$ and corresponding angular momentum $J(M_\mathrm{encl})$ by the dotted curves. The filled circles denote the points at which $j = j_\mathrm{ISCO}$ is satisfied for each progenitor model.}  
\label{fig:j_tff_M20}
\end{figure}

\begin{figure}
\centering
\includegraphics [width=0.48\textwidth]{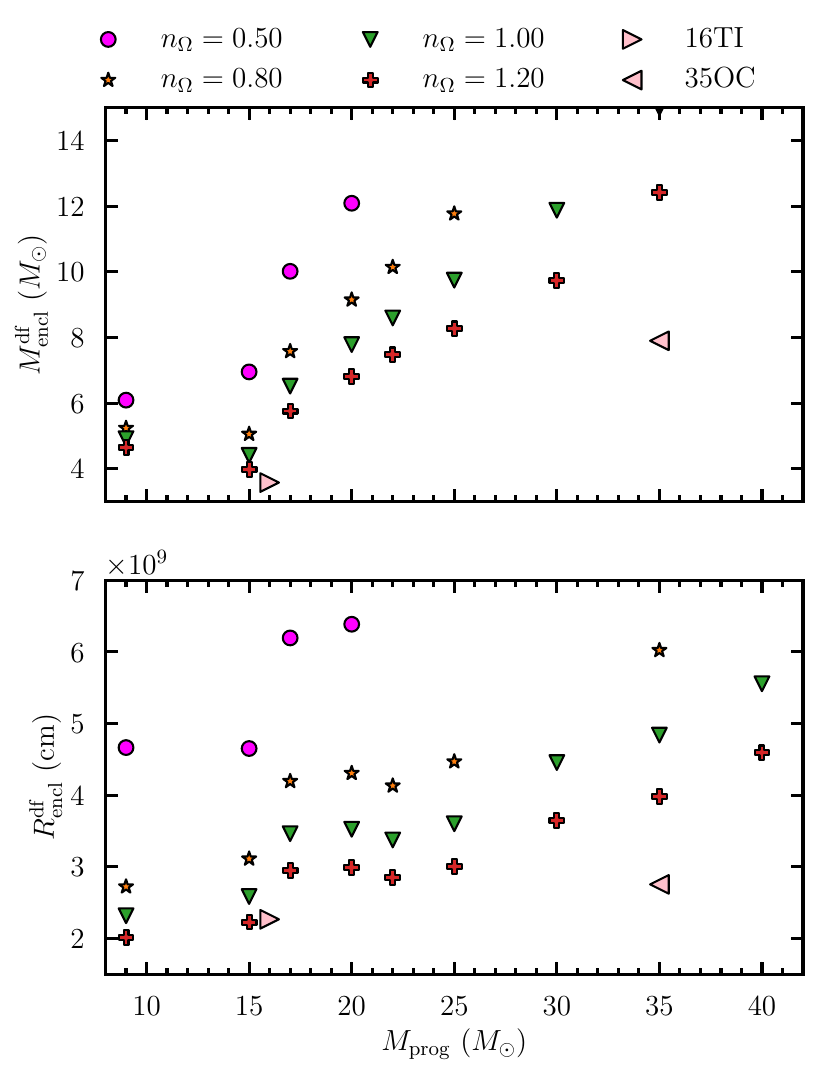}
\caption{Upper panel: estimated BH mass at the disc formation $M^\mathrm{df}_\mathrm{encl}$ as a function fo the progenitor mass $M_\mathrm{prog}$. $M^{\mathrm{df}}_\mathrm{encl}$ is evaluated to be the enclosed mass of the progenitor with a specific angular momentum equal to $j_\mathrm{ISCO}$. Lower panel: The radius $R^\mathrm{df}_\mathrm{encl}$ of the mass coordinate $M^\mathrm{df}_\mathrm{encl}$ as a function of $M_\mathrm{prog}$. The results are shown for a sample of the progenitor models from \citet{Aguilera-Dena_2020} with four degree of rotation $n_\Omega = 0.5, 0.8, 1.0, 1.2$ (results for progenitors with different magnitudes of the angular
velocity is distinguished by different markers). We also show the results for the models \texttt{16TI} and \texttt{35OC} from \citet{Woosley_2006} (pink right-pointing and left-pointing arrow respectively). }

\label{fig:M_prog_M_BH_R_BH}
\end{figure}

In Figure~\ref{fig:j_tff_M20}, we show the specific angular momentum of a mass shell, $(2/3)r^2\Omega$, as a function of $M_\mathrm{encl}$ for the model of $M_\mathrm{prog}=20\, M_\odot$ with different degrees of rotation: $n_\Omega = 0.5, 0.6, 0.8, 1.0$ and $1.2$ (solid lines, where the color distinguishes $n_\Omega$). We also plot $j_\mathrm{ISCO}$ for a BH that has mass $M_\mathrm{encl}$ and angular momentum $J_\mathrm{encl}$ for each model (dashed lines), and we highlight with filled circles the points at which $j = j_\mathrm{ISCO}$ before getting larger. We refer to the mass when $j = j_\mathrm{ISCO}$ as $M_\mathrm{encl}^{\mathrm{df}}$. When the mass shell falls into the center, a disc is expected to be formed. This plot shows that a BH is likely to grow more before the disc formation for lower values of $n_\Omega$, i.e., to $M_\mathrm{encl}^{\mathrm{df}} \approx 6.8, 7.8, 9.2, 10.1$ and $12.1$ $M_\odot$ for $n_\Omega = 1.2, 1.0, 0.8, 0.6$ and $0.5$, respectively.
This procedure of comparing the specific angular momentum with $j_\mathrm{ISCO}$ has applied to all the models to calculate the mass coordinate $M_\mathrm{encl}^{\mathrm{df}}$ and the corresponding radius $R_\mathrm{encl}^{\mathrm{df}}$. Figure~\ref{fig:M_prog_M_BH_R_BH} shows the results of this calculation as a function of the progenitor mass for both $M_\mathrm{encl}^{\mathrm{df}}$ (upper panel) and $R_\mathrm{encl}^{\mathrm{df}}$ (lower panel). In this plot, we present results for a sample of the models taken from \citet{Aguilera-Dena_2020} with $n_\Omega = 0.5, 0.6, 0.8, 1.0, 1.2$, and for the progenitor stars \texttt{16TI} and \texttt{35OC} (from \citet{Woosley_2006}, right-pointing and left-pointing triangle respectively).
The results presented in Figs.~\ref{fig:j_tff_M20} and \ref{fig:M_prog_M_BH_R_BH} confirm our hypothesis that in progenitor stars with lower angular velocity, the matter with sufficiently high angular momentum to form the disc is located at larger radii, hence presenting a longer disc formation time, $t_\mathrm{\mathrm{df}}$.

\begin{figure}
\centering
\includegraphics [width=0.48\textwidth]{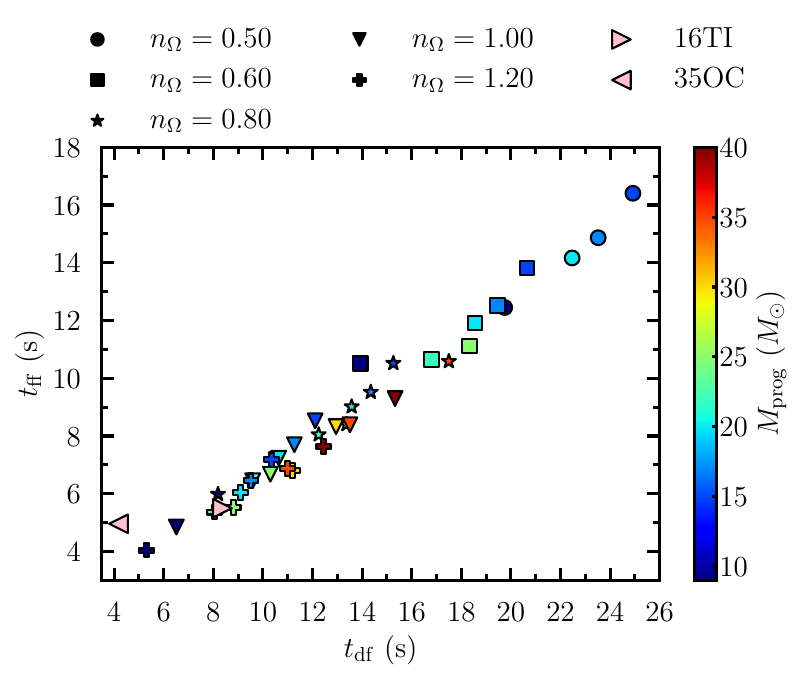}
\caption{Estimated free-fall time $t_\mathrm{ff}$ of the enclosed mass at which the disc is formed against the disc formation time $t_\mathrm{df}$ for a sample of the progenitor models from \citet{Aguilera-Dena_2020} with  four degrees of rotation: $n_\Omega = 0.5,0.8, 1.0,1.2$. Results for progenitors with different magnitude of the angular velocity is distinguished by different markers, while the color indicates the progenitor mass $M_\mathrm{prog}$. We also show the results for the models \texttt{16TI} and \texttt{35OC} from \citet{Woosley_2006} (pink right-pointing and left-pointing arrow respectively). }
\label{fig:M_prog_tff}
\end{figure}

We now calculate $t_\mathrm{ff}$ for the mass shell $M_\mathrm{encl}^{\mathrm{df}}$ as 
\begin{align}
t_\mathrm{ff} = \frac{\pi}{\sqrt{G M_\mathrm{encl}^{\mathrm{df}}}} \left(R_\mathrm{encl}^{\mathrm{df}}/2\right)^{3/2}. \label{t_ff}
\end{align}
In Figure~\ref{fig:M_prog_tff} we compare $t_\mathrm{ff}$ of the mass coordinate $M_\mathrm{encl}^{\mathrm{df}}$, at which the disc is expected to form (on the vertical axis), with $t_\mathrm{\mathrm{df}}$ measured in the simulations (on the horizontal axis) for the same sample of progenitor models from \citet{Aguilera-Dena_2020} used in Figure~\ref{fig:M_prog_M_BH_R_BH} (i.e., the progenitor stars with $n_\Omega= 0.5, 0.8, 1.0$ and $1.2$; $M_\mathrm{prog}$ is distinguished by color while the marker indicates $n_\Omega$) and the progenitor stars \texttt{16TI} and \texttt{35OC} (from \citet{Woosley_2006}, right-pointing and left-pointing triangle respectively). This plot clearly shows a linear correlation between the free-fall time analytically evaluated and the time at which the disc is formed in the simulations. It also confirms that the disc formation is supposed to take longer for slower rotating stars because $M_\mathrm{encl}^{\mathrm{df}}$ is located further out in the envelope. Focusing on the specific values of $t_\mathrm{ff}$ and $t_\mathrm{df}$, we notice that in our simulations, the disc formation time is by a factor of $\approx1.5$ longer than the estimation with the free-fall time. This is likely due to the fact that the stellar envelope is not really free-falling due to the presence of the pressure in our simulations.

The disc formation is employed as the condition triggering the generation of the wind in our simulations. The injection of matter and energy through the inner boundary does not begin simultaneously with the wind formation, though, since the wind needs some time to induce a positive mass flux at the inner boundary and make the matter there unbound. This happens once the ram pressure of the injected matter wins over that of the infalling envelope, which is lower in the later times. Both panels of Figure~\ref{fig:t_inj_t_df_Omega} show that, in our simulations, the time interval $t_\mathrm{inj}-t_\mathrm{df}$ varies from $\sim 2$~s for $n_\Omega=1.2$ to $\sim 10$~s for $n_\Omega=0.5$ and, for a fixed magnitude of the rotation, it remains mostly constant until $M_\mathrm{prog}\approx 25\, M_\odot$ and tends to slightly increase for $M_\mathrm{prog}> 30 \, M_\odot$. This can be explained considering that the stellar radius of the progenitors depends only weakly on the progenitor mass for the stellar-evolution models of \cite{Aguilera-Dena_2020}. Thus, for larger values of $M_\mathrm{prog}$, the ram pressure of the infalling matter can be higher, dominating over that of the injected matter, for longer times. 
Eventually, in all simulations, the ram pressure of the injected matter becomes larger than that of the infalling envelope, driving the explosion.

\begin{figure*}
\centering
	\includegraphics [width=0.48\textwidth]{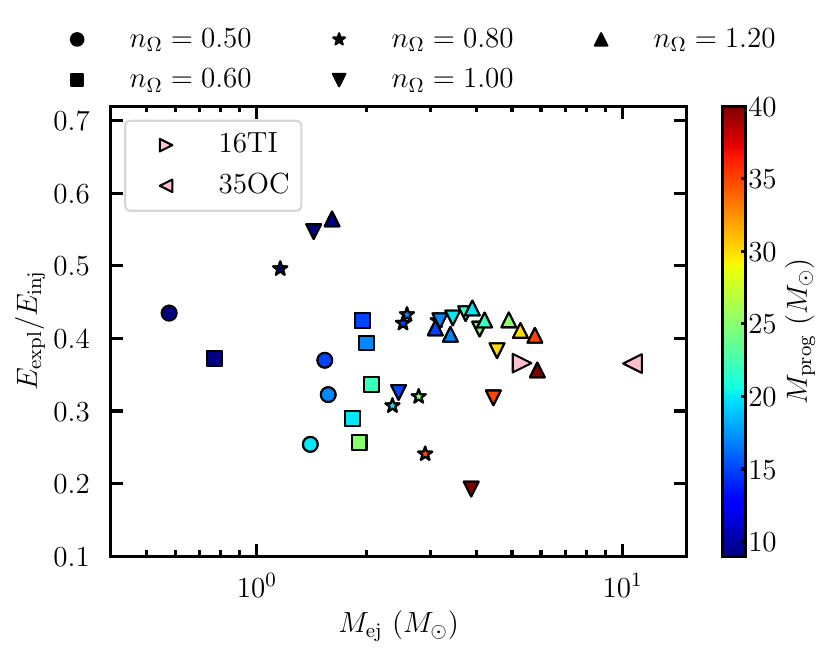}	
	\includegraphics [width=0.48\textwidth]{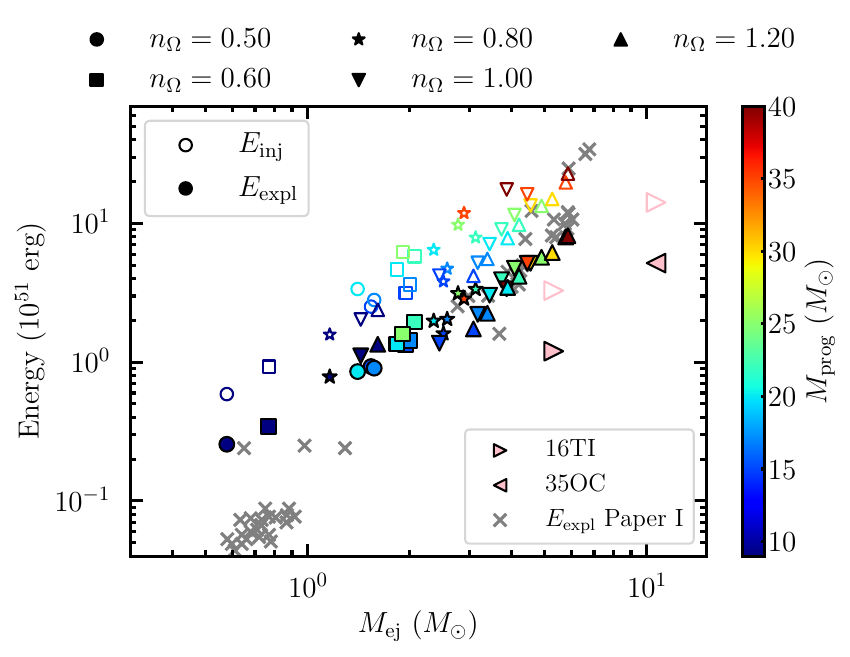}
\caption{Left panel: the ratio of the explosion energy to the injected energy $E_\mathrm{expl}/E_\mathrm{inj}$ as a function of the ejecta mass $M_\mathrm{ej}$ for all the models studied in this work. Right panel: $E_\mathrm{expl}$ (filled markers) and $E_\mathrm{inj}$ (open markers) as functions of the ejecta mass $M_\mathrm{ej}$. Results for progenitors from \citet{Aguilera-Dena_2020} with different degrees of rotation are distinguished by different markers, while the color indicates the progenitor mass $M_\mathrm{prog}$. The right-pointing and left-pointing pink triangles are the results for the models \texttt{16TI} and \texttt{35OC} from \citet{Woosley_2006}. The gray $\times$-markers show the results of $E_\mathrm{expl}$ obtained in \citetalias{LCMenegazzi} for the model AD020x1.0.}
\label{fig:Mej_Eexpl}
\end{figure*}

We then discuss the explosion energy and ejecta mass.
In Figure~\ref{fig:Mej_Eexpl}, we present the results of all simulations in terms of the injected and explosion energy ($E_\mathrm{inj}$  and $E_\mathrm{exp}$). In the left panel, we plot the ratio of the explosion energy to the injection energy, $E_\mathrm{expl}/E_\mathrm{inj}$, for the models from \citet{Aguilera-Dena_2020} with different values of $n_\Omega$ and from the models \texttt{16TI} and \texttt{35OC} as a function of $M_\mathrm{ej}$ and in the right panel, we display separately $E_\mathrm{inj}$ (open markers) and $E_\mathrm{expl}$ (filled markers) as functions of the ejecta mass. We also show the results of $E_\mathrm{expl}$ obtained in \citetalias{LCMenegazzi} for the model AD020x1.0 with gray $\times$-markers.

\begin{figure}
	\centering
	\includegraphics [width=0.48\textwidth]{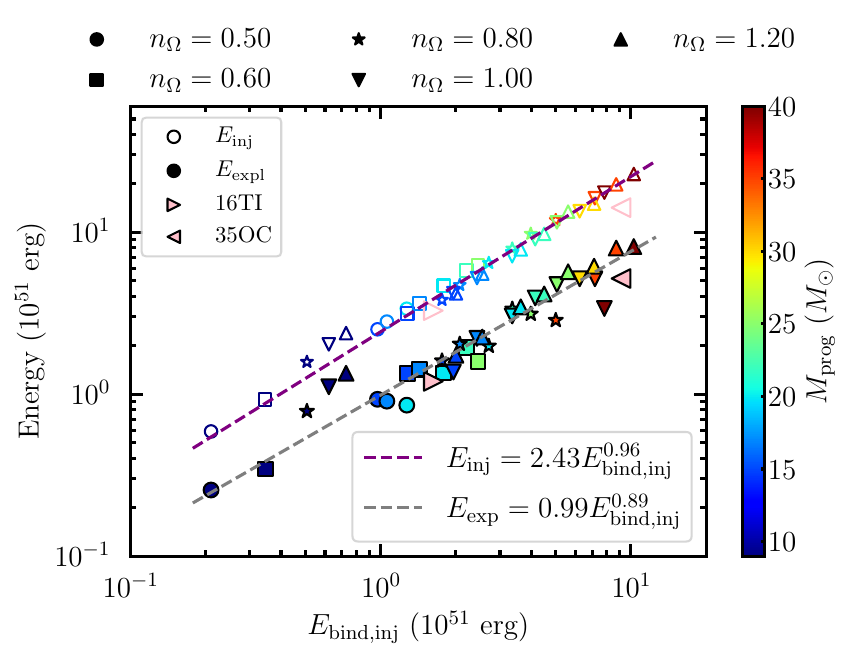}
	\caption{$E_\mathrm{expl}$ (filled markers) and $E_\mathrm{inj}$ (open markers) as functions of the binding energy evaluated at the injection time $E_\mathrm{bind,inj}$ for all the models studied in this work. The right-pointing and left-pointing pink triangles are the results for the models \texttt{16TI} and \texttt{35OC} from \citet{Woosley_2006}. The right-pointing and left-pointing pink triangles are the results for the models \texttt{16TI} and \texttt{35OC} from \citet{Woosley_2006}. The dashed purple line represents the linear regression $E_\mathrm{inj}$ and the dashed gray line the linear regression of $E_\mathrm{expl}$ for the models of \citet{Aguilera-Dena_2020}. The excluded  models are not shown in this plot and are discussed in Appendix~\ref{Appendix:exceptional_models}. Results for progenitors with different angular velocities are distinguished by different markers, while the color indicates the progenitor mass $M_\mathrm{prog}$.
	}
    \label{fig:Eexp_Einj_vs_Ebin}
\end{figure}

The left panel of Figure~\ref{fig:Mej_Eexpl} shows that the explosion energy represents 20--60\% of the injected energy for our models. From this plot, it is also evident that progenitors with higher angular velocity have a larger ejecta mass, while the values of $M_\mathrm{prog}$ seem to affect the efficiency of $E_\mathrm{expl}/E_\mathrm{inj}$, i.e., for more massive progenitors the ratio $E_\mathrm{expl}/E_\mathrm{inj}$ tends to be smaller. This behavior can be explained by considering that the more massive progenitors have more gravitational potential energy that the injected matter should overcome for the explosion (cf. Figure~\ref{fig:star_compactness}).

The right panel of Figure~\ref{fig:Mej_Eexpl} shows that more massive progenitors have larger values of $E_\mathrm{expl}$ and $M_\mathrm{ej}$ (in the case of the progenitors of \citet{Aguilera-Dena_2020}, for a fixed value of $n_\Omega$). Both $E_\mathrm{expl}$ and $E_\mathrm{inj}$ show a continuous distribution with respect to the ejecta mass. Considering the distributions of these quantities for the models of \citet{Aguilera-Dena_2020}, the explosion energy ranges from  $E_\mathrm{expl}$ = $0.3\times10^{51}$ erg for a model with $M_\mathrm{prog}= 9\; M_\odot$ and $n_\Omega =0.5$ to  $E_\mathrm{exp}$ = $8.2\times10^{51}$ erg for a model with $M_\mathrm{prog} = 40\; M_\odot$ and $n_\Omega =1.2$ and the injected energy from $E_\mathrm{inj}$= $0.6\times10^{51}$ erg to $E_\mathrm{inj}$ = $2.3\times 10^{52}$ erg for the same two models. The only point slightly detached from the rest of the distribution are those of the 9$M_\odot$ progenitor with $n_\Omega=0.5$ (AD009x0.5) and $n_\Omega=0.6$ (AD009x0.6). Nonetheless, the explosion energy follows the global trend even for these simulations. Therefore it is reasonable to assume that the space in the between would be filled by progenitors with initial mass ranging from 9 to 15 $M_\odot$ or $n_\Omega=0.7$. The distributions of $E_\mathrm{expl}$ and $E_\mathrm{inj}$  for the progenitor \texttt{16TI} and \texttt{35OC} also look continuous with respect to the ejecta mass and both the explosion and the injected energy are comparable to those measured for the models of \citet{Aguilera-Dena_2020} with analogous $M_\mathrm{prog}$, but the points are located at higher values of $M_\mathrm{ej}$. For the \texttt{16TI} we measure $E_\mathrm{expl}\approx 1.2\times 10^{51}$ erg and$M_\mathrm{ej}\approx 5.3\,M_\odot$ compared, for instance, to $E_\mathrm{expl}\approx 1.37\times 10^{51}$ erg and$M_\mathrm{ej}\approx 2.37\,M_\odot$ for AD015x1.5. For the \texttt{35OC} progenitor we instead measure $E_\mathrm{expl}\approx 5.2\times 10^{51}$ erg and $M_\mathrm{ej}\approx10.7\, M_\odot$ while in the case of AD035x1.0 we find $E_\mathrm{expl}\approx 5.2\times 10^{51}$ erg as well but $M_\mathrm{ej}\approx4.3\, M_\odot$.

The same figure also shows that $E_\mathrm{expl}$ and $M_\mathrm{ej}$ have a clear correlation with $n_\Omega$, i.e., the faster the progenitor rotates, the larger $E_\mathrm{expl}$ and $M_\mathrm{ej}$ are. This is consistent with our expectation since a progenitor with higher values of $n_\Omega$ has a larger disc mass which is the source of the wind injection.

It is also noticeable that the distribution of the points obtained in this work varying the progenitor model covers a smaller interval of explosion energies than that presented in \citetalias{LCMenegazzi} for different wind injection models that span from $\sim 5\times10^{49}$~erg to $\sim 3.4\times10^{52}$~erg. Additionally, in the present study, $E_\mathrm{expl}$ does not show a bimodal distribution for highly-energetic and sub-energetic explosions. This confirms our hypothesis presented in \S~\ref{sec:Results:dynamics} that the bimodal distribution obtained in \citetalias{LCMenegazzi} is likely to be determined by the model for central engine employed. Since for this analysis, we employ the parameters that determine an energetic explosion of AD020x1.0 in \citetalias{LCMenegazzi}, even changing the progenitor structure, the explosion would also belong to the same category. 

\begin{figure*}
	\centering
	\includegraphics [width=0.8\textwidth]{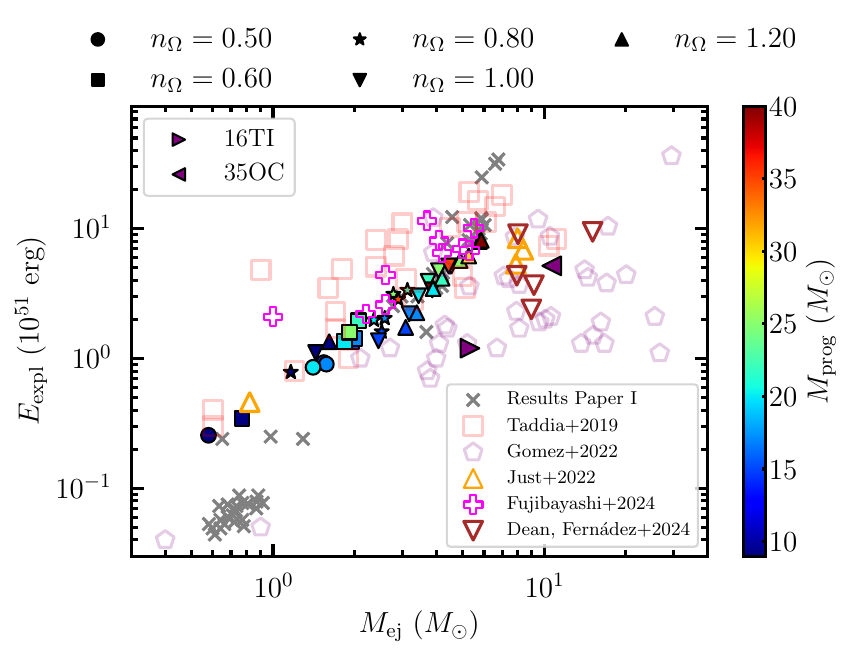} 
	\caption{Parameter dependence with respect to the observable pair of explosion energy $E_\mathrm{expl}$ and ejecta mass $M_\mathrm{ej}$. Results for progenitors from \citet{Aguilera-Dena_2020} with different degrees of rotation are distinguished by different markers, while the color indicates the progenitor mass $M_\mathrm{prog}$. The gray $\times$-markers show our results obtained in \citetalias{LCMenegazzi} for the model AD020x1.0. The open markers display the observational data for stripped-envelope SNe, some of which are Type Ic-BL SNe, taken from \citet{Taddia2019jan} and \citet{Gomez2022dec}. The up-pointing triangles display the results of \citet{Just2022} from neutrino-radiation viscous-hydrodynamics simulations with the progenitors \texttt{16TI}. The magenta plus-sign denotes the results obtained in a general relativistic neutrino-radiation viscous-hydrodynamics simulation for progenitors with $M_\mathrm{prog}=20,\,35,\,45\;M_\odot$ and $n_\Omega = 0.6,\,0.8,\,1.0,\,1.2$ \citep{Fujibayashi:2023oyt}. The down-pointing triangles show the results of \citet{Dean_2024} from neutrino-radiation viscous-hydrodynamics simulations with the progenitors \texttt{16TI} and \texttt{35OC}. For the same models  \texttt{16TI} and \texttt{35OC} we perform additional simulations displayed in the figure with a right-pointing and a left-pointing purple arrow respectively.
	}    \label{fig:Eexp_vs_Mej_w_obs_data}
\end{figure*}

In Figure~\ref{fig:Eexp_Einj_vs_Ebin}, we plot $E_\mathrm{inj}$ and $E_\mathrm{expl}$ as functions of the binding energy of the matter at the onset of the wind injection, $E_\mathrm{bind,inj}$, for all models. $E_\mathrm{inj}$ and $E_\mathrm{expl}$ increase approximately linearly with $E_\mathrm{bind,inj}$ and, specifically, the injected energy is always larger than the binding energy, as it is in the model 20\_3.16\_3.16\_0.1\_0.10. This confirms that in all the simulations, the explosion is driven by the same mechanism, i.e., by the ram pressure of the injected wind that dominates over that of the infalling matter, efficiently pushing forward the stellar envelope that expands without falling back. As a result, all the models experience an energetic explosion. 

A good linear correlation between $E_\mathrm{expl}$ and $E_\mathrm{bind,inj}$ is reasonable: the engine of the explosion has to provide the energy similar to $E_\mathrm{bind,inj}$ for a successful explosion. When the energy with the order of $E_\mathrm{bind,inj}$ is injected, the stellar envelope becomes unbound by the injected wind. Hence, the mass infall to the central BH-disc system and the new energy injection are suppressed. The explosion energy is thus likely to be regulated by the order of $E_\mathrm{bind,inj}$. Comparing the injected energy of the models \texttt{16TI} and \texttt{35OC} with that measured in for the progenitor of \citet{Aguilera-Dena_2020} for give $E_\mathrm{bind, inj}$, we notice that it is slightly smaller in the case of \texttt{16TI} and \texttt{35OC}. This difference can relate to the difference in the progenitor features and to the simple model we used. Therefore the fits shown in Figure~\ref{fig:Eexp_Einj_vs_Ebin} are made using only the results for the progenitors of \citet{Aguilera-Dena_2020}
.

\subsection{Comparison with the observations}\label{sec:Results:Comparison_obs}
In this section, we compare the ejecta properties obtained in our simulation with the observational data. Figure~\ref{fig:Eexp_vs_Mej_w_obs_data} presents the distribution of our model in the $E_\mathrm{expl}$-$M_\mathrm{ej}$ plane (filled markers) along with the observational data for Ic-BL SNe taken from \citet{Taddia2019jan} and for stripped-envelope SNe, including Type Ic-BL SNe from \citet{Gomez2022dec} (open markers). 
We additionally display the results of some general relativistic neutrino-radiation viscous-hydrodynamics simulations obtained using different progenitors: from \citet{Fujibayashi:2023oyt} for progenitors with $M_\mathrm{prog}=$20, 35 and 45 $M_\odot$ of \citet{Aguilera-Dena_2020} and different degrees of rotation $n_\Omega = 0.6,\,0.8,\,1.0,\,1.2$ and from \citet{Just2022, Dean_2024} for \texttt{16TI} and \texttt{35OC} models in \citet{Woosley_2006}.  
In Figure~\ref{fig:Eexp_vs_Mej_w_obs_data}, we focus on the dependence of the explosion on the progenitor mass $M_\mathrm{prog}$ indicated by the color of the markers and on the magnitude of the angular velocity $n_\Omega$ distinguished by the marker shape.

Comparing our numerical results with the observational data, we find that our results agree with some range of the observational data. However, we also find that, despite the wide-ranging variations of $M_\mathrm{prog}$ and $n_\Omega$, the explosion energy distributes along a trend and does not show the extended distribution made by observational data. In other words, the explosion energy and the ejecta mass are more strongly correlated in our simulations than in the observational data provided by \citet{Taddia2019jan} and \citet{Gomez2022dec}. A similar tight correlation between $E_\mathrm{expl}$ and $M_\mathrm{ej}$ was also found in our previous work for the model AD020x1.0 as indicated by the $\times$-markers in the right panel of Figure~\ref{fig:Eexp_vs_Mej_w_obs_data} (see also Figure~11 in \citetalias{LCMenegazzi}).

We find that the observational data points with $M_\mathrm{ej}\gtrsim10M_\odot$ are hardly reproduced with the progenitor models of \citet{Aguilera-Dena_2020}. The reason is just that the available mass for the ejecta, i.e., the pre-collapse stellar mass minus BH mass at the disc formation (see the upper panel of Figure~\ref{fig:M_prog_M_BH_R_BH} and upper panel of Figure~1 in \citealt{Aguilera-Dena_2020}), is at most $\sim10M_\odot$. The events with $M_\mathrm{ej}\gtrsim10M_\odot$ may have originated from the stars that have larger available mass, i.e., with larger envelope mass or faster rotation, in the context of a collapsar scenario.

Our speculation is supported by the points obtained by numerical simulations (\citealt{Dean_2024,Just2022}) that use faster rotating pre-collapse structure (\texttt{16IT} and \texttt{35OC}; \citealt{Woosley_2006}). Their ejecta mass is systematically higher than those of our result. This indicates that the tight correlation found in our result may stem from our limited choice of the progenitor structure, and we could reproduce events with $M_\mathrm{ej}\gtrsim10M_\odot$ with more massive or faster-rotating progenitors.

\begin{figure}
	\centering
	\includegraphics [width=0.48\textwidth]{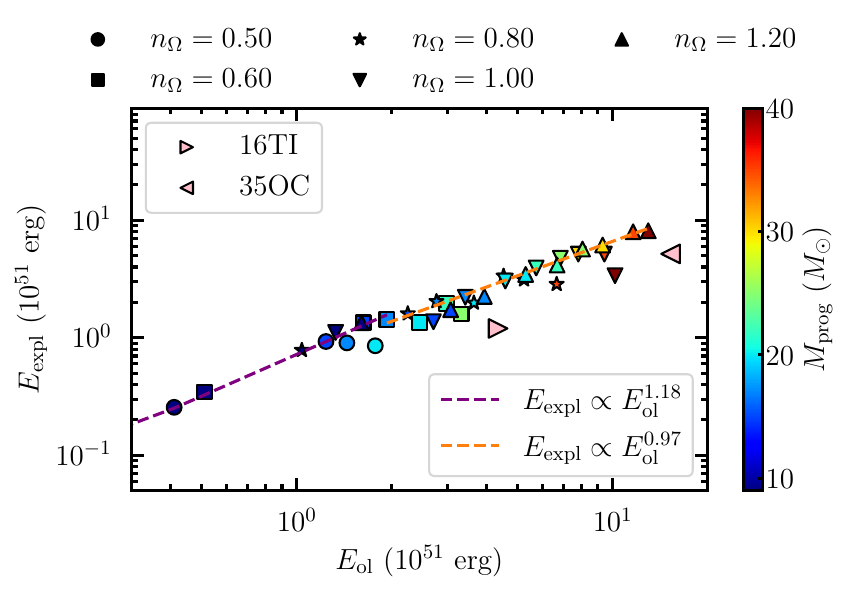}
	\caption{Explosion energy $E_\mathrm{expl}$ as a function of the binding energy of the outer layers in the pre-collapse ($E_\mathrm{ol}$) for the models from \citet{Aguilera-Dena_2020}. $E_\mathrm{ol}$ is a measure of the compactness of the entire star.
    The dashed purple line shows the best fit obtained with a broken power-law.
	Results for progenitors with different angular velocity are distinguished by different markers, while the color indicates the progenitor mass $M_\mathrm{prog}$. The right-pointing and left-pointing pink triangles are the results for the models \texttt{16TI} and \texttt{35OC} from \citet{Woosley_2006}. }
    \label{fig:Eexp_compactness_Binding_energy}
\end{figure}

We also note that the results obtained in \citet{Fujibayashi:2023oyt} are well aligned with the outcomes of our simulations showcased in Figure~\ref{fig:Eexp_vs_Mej_w_obs_data}, while the results in \cite{Dean_2024} and \cite{Just2022} are located in the regime with systematically higher values of $M_\mathrm{ej}$. The better agreement with \citet{Fujibayashi:2023oyt} could also be due to the choice of the progenitor models: \citet{Fujibayashi:2023oyt} also use the same progenitor stars as ours. This fact may also support the above speculation.

To confirm our speculation, we perform additional simulations using progenitor models \texttt{16TI} and \texttt{35OC}. For the \texttt{16TI} we measure the ejecta mass of $M_\mathrm{ej}\approx 5.3\,M_\odot$ and en explosion energy $E_\mathrm{expl}\approx 1.2\times 10^{51}$ erg, while for the \texttt{35OC} progenitor $M_\mathrm{ej}\approx10.7\, M_\odot$ and $E_\mathrm{expl}\approx 5.2\times 10^{51}$ erg.
For these models we measure an $E_\mathrm{expl}$ similar to that of the progenitor from \citet{Aguilera-Dena_2020} with the same mass, but the point distribution \texttt{16TI} and \texttt{35OC} is located at higher $M_\mathrm{ej}$, in qualitative agreement with the results obtained by \citet{Dean_2024}. 
The difference between the values of $E_\mathrm{expl}$ we measured for \texttt{16TI} and \texttt{35OC} and those obtained by \texttt{16TI} and \texttt{35OC} can be attributed to the simple wind model we used where the wind parameters are fixed throughout the whole simulation.

\subsection{Prediction on the Explosion Energy}\label{sec:Results:explodability_prediction}

It would be advantageous to have a tool useful for choosing a potential progenitor based on some requirements for the outcomes before performing any simulation. We provide such a tool using the binding energy of the outer layers $E_\mathrm{ol}$ ($\mathrm{ol}$ = outer layers), which is defined as:
\begin{align}\label{compactness_our}
    E_\mathrm{ol}=\frac{G M_\mathrm{ol}M_\mathrm{encl}^{\mathrm{df}}}{R_\mathrm{encl}^{\mathrm{df}}}\;,
\end{align}
where $M_\mathrm{encl}^{\mathrm{df}}$ is the mass coordinate at which the specific angular momentum is equal to that of ISCO for a BH with the same mass and angular momentum of their enclosed values, i.e., the estimated BH mass at the disc formation. 
$R_\mathrm{encl}^{\mathrm{df}}$ is the radius of the mass coordinate $M_\mathrm{encl}^{\mathrm{df}}$, and $M_\mathrm{ol}$ is the mass outside the same mass coordinate. This quantity would provide an estimate of available energy by the accretion of the outer layer to the disc.

In Figure~\ref{fig:Eexp_compactness_Binding_energy} we show the relation between $E_\mathrm{expl}$ and $E_\mathrm{ol}$, for both the model from \citet{Aguilera-Dena_2020} and the \texttt{16TI} and \texttt{35OC} of \citet{Woosley_2006} (right and left-pointing pink triangles respectively). Focusing on the models of \citet{Aguilera-Dena_2020}, we find that these quantities have a correlation that can be expressed reasonably by
 \begin{align}\label{E_expl_E_ol_bpl}
E_\mathrm{expl} \approx \SI{1e51}{erg}\times 
\begin{cases}
\Bigg(\frac{E_\mathrm{ol}}{\SI{2e51}{erg}}\Bigg)^{1.18} & (E_\mathrm{ol}<2\times10^{51}\,\mathrm{erg}), \\
\Bigg(\frac{E_\mathrm{ol}}{\SI{2e51}{erg}}\Bigg)^{0.96} & (E_\mathrm{ol}>2\times10^{51}\,\mathrm{erg}). \\
\end{cases}
\end{align}

The above equation provides a prediction for the explosion energy only from the pre-collapse conditions. The approximate linear dependence of the explosion energy on $E_\mathrm{ol}$ for $E_\mathrm{ol}>\SI{2e51}{erg}$ may stem from the same reason as its dependence on $E_\mathrm{bind,inj}$ (see Figure~\ref{fig:Eexp_Einj_vs_Ebin} in \S~\ref{sec:Results:dynamics}). Its steeper dependence for $E_\mathrm{ol}<\SI{2e51}{erg}$ might relate to progenitor star, more specifically to the specific gravitational binding energy at the surface, $e_\mathrm{bind,surface}$, of the pre-collapse progenitor models (see Figure~\ref{fig:star_compactness} in \S~\ref{sec:Progenitor_Parameters}). $e_\mathrm{bind,surface}$ shows in Figure~\ref{fig:star_compactness} a monotonic increase with the progenitor mass that is steeper for $M_\mathrm{prog}<25\, M_\odot$. This may lead to a smaller energy injection necessary for making the stellar envelope unbound, resulting in a smaller explosion energy compared to $E_\mathrm{ol}$ for $E_\mathrm{ol}<\SI{2e51}{erg}$.  

Considering the relation between $E_\mathrm{expl}$ and $E_\mathrm{ol}$ for the progenitor stars \texttt{16TI} and \texttt{35OC}, we notice that the points lay outside the trend described by the distribution of the progenitors from \citet{Aguilera-Dena_2020}. Therefore the relation we found between $E_\mathrm{ol}$ and $E_\mathrm{expl}$ applies only to the models of \citet{Aguilera-Dena_2020} with a moderate modification of rotation.

\subsection{$^{56}$Ni production}\label{sec:Results:Ni_production}
In this section, we analyze the results of the $^{56}$Ni production and the way in which it correlates with the progenitor mass and its angular velocity. By doing that, we aim to ascertain whether it is feasible to replicate observational data such as those presented by \citet{Taddia2019jan} and \citet{Gomez2022dec}, especially for the high-energy SNe ($E_\mathrm{expl} \gtrsim 10^{52}$ erg). 

\begin{figure*}
\centering
	\includegraphics [width=0.48\textwidth]{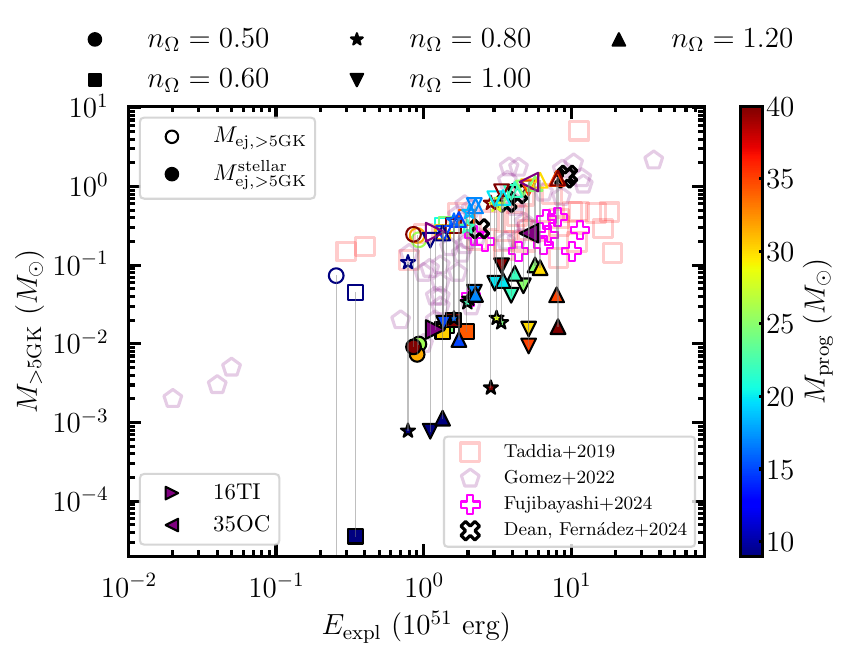}	
	\includegraphics [width=0.48\textwidth]{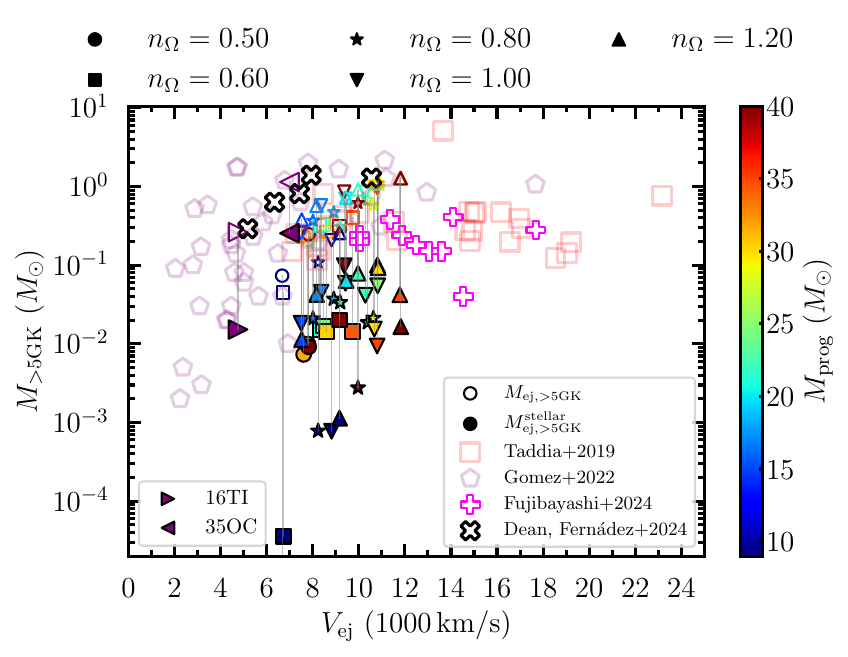}
\caption{Relations between the explosion energy and the $^{56}$Ni mass (left) and between the average velocity of the ejecta and the $^{56}$Ni mass (right). Each gray line connects $M^\mathrm{stellar}_\mathrm{ej,>5GK}$ (triangles) and $M_\mathrm{ej,>5GK}=M^\mathrm{stellar}_\mathrm{ej,>5GK} + M^\mathrm{inj}_\mathrm{ej,>5GK}*M^\mathrm{inj}_\mathrm{>5GK}/M^\mathrm{inj}$ (down-pointing triangles) of the same model to show the possible range of $^{56}$Ni mass. For the models from \citet{Aguilera-Dena_2020}, results for progenitors with different angular velocity are distinguished by different markers, while the color indicates the progenitor mass $M_\mathrm{prog}$. The results with the models \texttt{16TI} and \texttt{35OC} are displayed  with a right-pointing and a left-pointing purple arrow respectively. The open markers display the observational data for stripped-envelope SNe, some of which are Type Ic-BL SNe, taken from \citet{Taddia2019jan} and \citet{Gomez2022dec}. The magenta plus-sign denotes the results obtained in a general relativistic neutrino-radiation viscous-hydrodynamics simulation for progenitors with $M_\mathrm{prog}=20,\,35,\,45\;M_\odot$ and $n_\Omega = 0.6,\,0.8,\,1.0,\,1.2$ \citep{Fujibayashi:2023oyt}. The open black x-markers show the results of \citet{Dean_2024} from neutrino-radiation viscous-hydrodynamics simulations with the progenitors \texttt{16TI} and \texttt{35OC}.
}
\label{fig:M_Ni_w_obs_data}
\end{figure*}

The data used in our $^{56}$Ni calculations are summarized in Table~\ref{tab:models}. The table includes, starting from the fifth column: the mass of ejecta component originating from the injected matter (i.e., from the disc) $M^\mathrm{inj}_\mathrm{ej}$, the total mass experiencing temperature higher than 5~GK, $M_\mathrm{>5GK}$, (5~GK is the threshold above which the $^{56}$Ni production primarily occurs), the mass of the stellar component experiencing temperature higher than 5~GK, $M^\mathrm{stellar}_\mathrm{ej,>5GK}$, and the ratio between the injected matter that is estimated to experience temperature higher than 5~GK and the
total injected mass $M^\mathrm{inj}_\mathrm{>5GK}/M^\mathrm{inj}$. For the values of $M_\mathrm{>5GK}$ and $M^\mathrm{stellar}_\mathrm{ej,>5GK}$, also the number of tracers is shown in parenthesis. 
In this work, we employ tracer particles to estimate the $^{56}$Ni mass synthesized in the matter originating from the stellar envelope under the assumption that it is produced by the fluid elements experiencing temperatures higher than 5~GK.
We make this approximation without computing the whole nucleosynthesis because we found that the ejecta mass is dominated by the injected matter, which is larger than $M^\mathrm{stellar}_\mathrm{ej,>5GK}$ by more than one order of magnitude. For this component, as mentioned in \S~\ref{sec:Method} we estimate the mass of $^{56}$Ni by evaluating the temperature of the disc when the matter is injected (see Appendix~\ref{Appendix:T_inj_matter}).
Therefore, we roughly estimate the mass of $^{56}$Ni in the ejecta, $M_\mathrm{ej,Ni}$ as  $M_\mathrm{ej,Ni} = M^\mathrm{stellar}_\mathrm{ej,>5GK}+M^\mathrm{inj}_\mathrm{ej,>5GK}$, where $M^\mathrm{inj}_\mathrm{ej,>5GK}$ is $M^\mathrm{inj}_\mathrm{ej}$ multiplied by the ratio $M^\mathrm{inj}_\mathrm{>5GK}/M^\mathrm{inj}$.

With this approximation, $M_\mathrm{ej,Ni}$ is found to represent the $\sim 13$--$41\%$ of the total ejecta mass, with the lowest amount of $\sim0.19\,M_\odot$ produced in the progenitor AD009x0.8 and the largest of $\sim2.3\, M_\odot$ for the model AD040x1.2.

Figure~\ref{fig:M_Ni_w_obs_data} shows for all models our estimates of the $^{56}$Ni mass produced in the whole ejecta $M_\mathrm{ej,>5GK}$ (open markers) and that originating from the stellar component $M^\mathrm{stellar}_\mathrm{ej,>5GK}$ (filled markers) as a function of the explosion energy (left panel) and of the average ejecta velocity (right panel). In addition to the results of our simulations, in the plots we include the observational data for Type Ic-BL SNe taken from \citet{Taddia2019jan} and for stripped-envelope SNe, including Ic-BL SNe, taken from \citet{Gomez2022dec}. Furthermore we also show the $^{56}$Ni mass obtained by \citet{Fujibayashi:2023oyt} and by \citet{Dean_2024}.

Figure~\ref{fig:M_Ni_w_obs_data} highlights how strongly the component of the ejecta originating from the injected matter dominates the estimate of the total $^{56}$Ni mass produced in all models. It also shows that $M_\mathrm{ej,>5GK}$ tends to increase with $E_\mathrm{expl}$ (or with respect to $v_\mathrm{ej}$). The dependence of $M_\mathrm{ej,>5GK}$ on $E_\mathrm{expl}$ is reasonable: more massive and faster-rotating progenitors have larger values of $E_\mathrm{expl}$ and $M_\mathrm{ej}$ due to the larger disc mass (we discussed it in \S~\ref{sec:Results:dynamics}; see also Figure~\ref{fig:Mej_Eexpl}) and $M_\mathrm{disc}$ is the source of the injected matter, a significant fraction of which is here considered to become $^{56}$Ni. Even though not as linearly as $M_\mathrm{ej,>5GK}$, also  $M^\mathrm{stellar}_\mathrm{ej,>5GK}$ roughly increases with $E_\mathrm{expl}$ in the left panel of Figure~\ref{fig:M_Ni_w_obs_data}. This is because the energy injection occurs in similar timescale $\sim t_\mathrm{w}$. This leads to the higher energy injection rate $E_\mathrm{inj}/t_\mathrm{w}$ for higher $E_\mathrm{expl}$ models, and thus, more matter tends to experience higher temperature (see \citealt{Suwa2019} for the positive correlation between the energy injection rate and temperature that the ejecta experience).

The left panel of Figure~\ref{fig:M_Ni_w_obs_data} shows that our numerical estimates of the $^{56}$Ni mass are in fair agreement with the relation between $M_\mathrm{ej,Ni}$ and $E_\mathrm{expl}$ in the observational data and with the results obtained by both \citet{Fujibayashi:2023oyt} and \citet{Dean_2024}. We can reproduce at least some class of SNe Ic-BL with, basically, only the exclusion of the observed explosions with $E_\mathrm{expl}<10^{50}$ erg or $E_\mathrm{expl}>10^{52}$ erg. The capability of reproducing the observational data is the result of the calibration of the wind parameter with the numerical results of \citet{Fujibayashi:2023oyt}. Figure~\ref{fig:M_Ni_w_obs_data} shows that such calibration allows us to be in good agreement with a wide range of energetic SNe with a broad range of progenitor mass.
Comparing our results with the observational data in the right panel of Figure~\ref{fig:M_Ni_w_obs_data}, we also have to note that lower-velocity SNe of \citet{Gomez2022dec} are likely to be normal SNe Ic resulting from the heating by neutrinos emitted from the PNS, which we are not trying to reproduce with our models.

Focusing on the right panel of Figure~\ref{fig:M_Ni_w_obs_data}, we observe that  
the total $^{56}$Ni mass produced in our simulations is in fair agreement with the observational data in the range of the ejecta velocity we measured, i.e., $v_\mathrm{ej}\sim 6\times10^3$--$12 \times10^3$ km/s. However, our simulation results are less scattered than the observational data, which spread up to $\sim 25\times 10^3$ km/s (excluding the normal SNe Ic from \citet{Gomez2022dec} and focusing on the high-energy distribution). 
Considering the fact that the results of \cite{Fujibayashi:2023oyt} extend to higher velocity up to \SI{18e3}{km/s}, the reasonable reason for this can be attributed to our choice of a simple wind model; for instance, in the present setup, the wind time scale $t_\mathrm{w}$ and the accretion time scale $t_\mathrm{acc}$ are set constant and do not depend on the BH-disc properties, e.g., the Keplerian time at the typical disc radius $\propto \sqrt{M_\mathrm{BH}/ {r_\mathrm{disc}}^3}$.
We find that the higher progenitor mass tends to result in high ejecta velocity in \cite{Fujibayashi:2023oyt}. This indicates that our wind parameter set chosen in this study is somewhat different for higher progenitor mass. Considering that \citetalias{LCMenegazzi} achieved a high ejecta velocity $\sim \SI{20e3}{km/s}$ with longer wind timescale $t_\mathrm{w}=\SI{10}{s}$ for $M_\mathrm{prog}=20M_\odot$ star, the disc formed in the collapse of a more massive progenitor may have a longer wind timescale.

\begin{figure}
\centering
\includegraphics [width=0.48\textwidth]{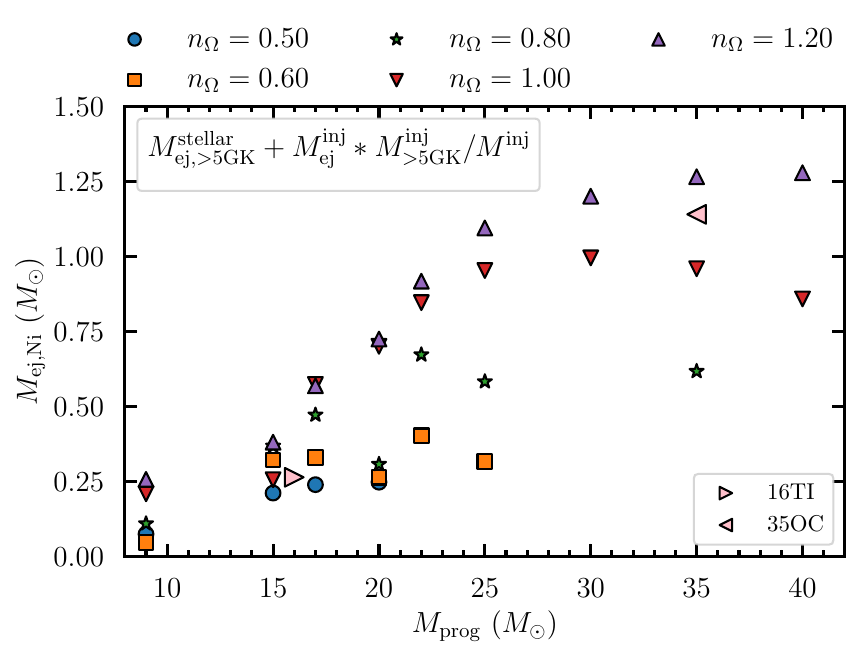}	
\caption{Correlation between the $^{56}$Ni mass and the progenitor mass with a variety of progenitor angular velocity. For the progenitors of \citet{Aguilera-Dena_2020}, markers and colors distinguish the values of $n_\Omega$. The results with the models \texttt{16TI} and \texttt{35OC} are displayed  with a pink right-pointing and a left-pointing purple arrow respectively.
} 
\label{fig:M_Ni_M_prog}
\end{figure}

In Figure~\ref{fig:M_Ni_w_obs_data}, it is possible to observe a dependence of the total $^{56}$Ni mass in the ejecta on the progenitor mass and angular velocity. To better appreciate it, in Figure~\ref{fig:M_Ni_M_prog}, we plot $M_\mathrm{ej,Ni}$ as a function of $M_\mathrm{prog}$ for all models, distinguishing them also by their degree of rotation with different markers and colors. The figure shows a specific correlation among $M_\mathrm{ej,Ni}$, $M_\mathrm{prog}$ and $n_\Omega$. More massive and faster-rotating progenitors tend to produce more $^{56}$Ni in the ejecta. The explanation for this behavior can be the same as speculated above to describe the relation between $M_\mathrm{ej,Ni}$ and $E_\mathrm{expl}$: more massive and faster-rotating progenitors have a larger disc mass, which is the source of the wind injection and $^{56}$Ni mass.


It is worth reminding that in this work, we assume that the entire stellar component of the injected matter experiencing $T>5$~GK becomes $^{56}$Ni. However, if the ejecta electron fraction is lower (i.e., $Y_e < 0.5$), the estimates of the $^{56}$Ni would be different. In such a scenario, the nucleosynthesis yield is not expected to peak at $^{56}$Ni, but rather at heavier nuclei (\citealt{Siegel_2019Natur}). Moreover, the amount of $^{56}$Ni generated in the injected component should not be substantial. Consequently, the mass of $^{56}$Ni would not be expected to be mainly determined by $M_\mathrm{ej}^\mathrm{inj}$.

\section{Discussion}\label{sec:Discussion}

\subsection{Variety of disc wind-driven explosion }

The present analysis of our results shows a large variety of ejecta mass with $M_\mathrm{ej}$ going from $\sim 0.6$ $M_\odot$ for the model AD009x0.5 to $>10$ $M_\odot$ for the progenitor \texttt{35OC} and a large variety of explosion energy with $E_\mathrm{expl}$ spanning from $\sim 0.3\times10^{51}$ erg for the model AD009x0.5 to $\sim 8\times10^{51}$ erg for AD040x1.2.  Focusing on the results obtained using the models of \citet{Aguilera-Dena_2020}, Figure~\ref{fig:Mej_Eexpl} shows a monotonic trend of our results, where $E_\mathrm{expl}$ and $M_\mathrm{ej}$ increase with the progenitor mass and the initial degree of the rotation. The effect of the progenitor mass on the outcome of the explosion can be explained considering that, typically, more massive stars are supposed to have a more compact envelope (in terms of $M_*/R_*$), resulting in higher mass-infall rates, which provides a larger amount of matter for the disc formation and a larger energy budget for explosion energy. The impact of the initial magnitude of the rotation, after the BH formation, on the explosion energy, and on the ejecta mass is investigated by fully general relativistic radiation viscous hydrodynamics in \citet{Fujibayashi:2023oyt}. 
They showed that a star with a fast rotation can yield a more energetic explosion and enhance mass ejection. Our present models approximately reproduce their findings.

Another interesting result, as mentioned in Section~\ref{sec:Results:Comparison_obs}, is that we found a stronger correlation between $E_\mathrm{expl}$ and $M_\mathrm{ej}$ in our models than in the observational data. We also hardly reproduced data points with $M_\mathrm{ej}\gtrsim10 \, M_\odot$ (refer to Figure~\ref{fig:Eexp_vs_Mej_w_obs_data}). As discussed in \S~\ref{sec:Results:Comparison_obs}, they may be partially due to the choice of the progenitor models. In the same subsection, we demonstrated that using the faster-rotating pre-collapse structures \texttt{16TI} and \texttt{35OC}, also employed by \citet{Dean_2024}, we could actually reproduce samples with $M_\mathrm{ej}\gtrsim10 \, M_\odot$. This indicates that there may be a wider variety of ejecta properties than we find in this work, which stems from the difference in the mass and angular momentum distributions of the pre-collapse stellar structure.

Another possible reason for the tight correlation can be attributed to the simple wind model we use, in which the wind parameters, e.g., the wind and accretion time-scales, are fixed throughout the whole simulation. As mentioned in \S~\ref{sec:Progenitor_Parameters}, this hypothesis is supported by the comparison to the results obtained in \cite{Fujibayashi:2023oyt}. The higher mass progenitor models tend to result in higher velocities and, thus, higher explosion energies than we find with the same progenitors (see also Figure~\ref{fig:Mej_Eexpl}). This indicates that the relevant timescale for the wind injection may be different for different progenitors. This is reasonable because the different mass and angular momentum distribution of pre-collapse structure naturally leads to the different characteristics of the BH-disc system, e.g., different Keplerian timescale.
We would, then, conclude that it may be possible to reproduce the variety of the observational data by the combination of the wide variety of the progenitor structure and more consistent modeling of the wind injection. 
In reality, the wind injection would occur after the efficiency of neutrino cooling compared to the viscous heating in the disc drops \citep{Fujibayashi:2023oyt}; both $t_\mathrm{w}$ and $t_\mathrm{acc}$ should depend sensitively on the neutrino cooling.
Therefore, it is necessary to construct a more sophisticated wind injection model that consistently captures the physical processes during both the neutrino-dominated accretion flow (NDAF; \citealt{Popham_Woosley_1999}, \citealt{Kohri_2005}) and the advection-dominated accretion flow (ADAF; \citealt{Narayan_Yi_1994} and \citealt{Hayakawa_2018}) phases in the specific case of a viscosity-driven wind explosion. We leave the construction of the model and the investigation with it for future work.

\subsection{The effect of GRB jet}
One of our aims for future studies and a possible use of this work is to connect the progenitor and the failed CCSN with a GRB that could be launched in such a scenario if a relativistic jet is produced (see, e.g., \citealt{Aloy_2000, Izzard_Robert_Tout, Zhang_2004, Mizuta_2006, Gottlieb_Ore_Lalakos, Shibata:2023tho} for simulation works with various sophistication). 
Performing relativistic-hydrodynamic simulations with the inclusion of relativistic jets would be an interesting case to investigate. As a matter of fact, if we also consider in our model a large dimensionless spin of the BH  and an electromagnetic fields, then we could have the formation of an energetic jet or outflow along the spin axis of the BH determined by the Blandford-Znajek effect (see \citealt{Blandford_Znajek}). Consequently, if a relativistic jet is formed, the energy budget for the explosion and the $^{56}$Ni production may increase because more energy is injected into the stellar matter. 
Several works have already studied this scenario, hence proposing that the jet and associated cocoon drive the stellar explosion and that it can launch a GRB (e.g., \citealt{Hjorth_2003, Stanek_2003, Lazzati_2012, Suzuki:2021ceq, Eisenberg:2022law}). \citet{Tominaga_2007b} and \citet{Tominaga2008aspherical} studied the jet-induced explosions of a Population III $40\;M_\odot$ star and suggested a correlation between GRBs with and without bright SNe and the energy deposition rate $\dot{E}_\mathrm{dep}$ (see also \citealt{Maeda_2003, Nagataki_2006}). They found that explosion with high energy ($\dot{E}_\mathrm{dep}\gtrsim 6\times 10^{52}$ erg) can synthesize a large amount of $^{56}$Ni ($\gtrsim  0.1\;M_\odot$) resulting consistent with GRB-SNe.  Contrarily, if the explosion deposition rate is low ($\dot{E}_\mathrm{dep}\lesssim 3\times 10^{51}$ erg), they measured a low ejected $^{56}$Ni mass ($\lesssim  10^{-3}\;M_\odot$) comparable to that observed in GRBs without SN brightening (like GRB060505 and GRB0606014). The GRB-SN mechanism was also found to be strongly sensitive to the angular momentum of the progenitor, but unaffected by its mass \citep{Hayakawa_2018}. 
Considering the previous works, then, one of our follow-up studies will be performing relativistic-hydrodynamic simulations that incorporate the injection of relativistic jets.

\section{Summary and conclusions}\label{sec:Conclusion}
In this work, we extended our previous study of the hydrodynamics and nucleosynthesis for the explosion of massive stars in the collapsar scenario by performing a series of two-dimensional, Newtonian simulations of progenitor models with different characteristics. Specifically, we selected nine models from \citet{Aguilera-Dena_2020} sampled in the range of $M_\mathrm{prog}= 9$--$40 \; M_\odot$.  We, then, studied them at five different angular velocities (see \S~\ref{sec:Method}). To further investigate the dependence of the ejecta properties on the progenitor structure, we additionally performed two simulations with the models \texttt{16TI} and \texttt{35OC} from \citet{Woosley_2006} that have a larger angular momentum than those of \citet{Aguilera-Dena_2020} for a given mass (for these simulations we used the original angular velocity profile).
For these simulations, we worked with the open-source multi-dimensional hydrodynamics code \texttt{Athena++} solving the axisymmetric gravitational potential as in \citetalias{LCMenegazzi}. We also used the same model for the central engine that is supposed to evolve the BH and the disc with the transfer of matter and angular momentum and that also includes the disc wind formation and injection.

Our main aim was to investigate the effect of the progenitor structure on the properties of the ejecta.  
In order to focus our investigation on the effect of the progenitor structure on the faith of the evolution, we fix the parameters of the wind injection model so that the results of the explosion of the AD020x1.0 progenitor are similar to those obtained by \citet{Fujibayashi:2022xsm} for the same star. 
In all our models, the disc-driven explosion results in an explosion with $E_\mathrm{expl}$ ranging from $0.3\times10^{51}$~erg for the model AD009x0.5 to $>8\times 10^{51}$~erg for the model AD040x1.2 for the progenitor models of \citet{Aguilera-Dena_2020}. For the progenitor models of \citet{Woosley_2006}, the explosion energy is comparable for a given progenitor mass, but the ejecta mass is larger because they have a larger available mass (the pre-collapse stellar mass minus BH mass at the disc formation), determined by a faster rotation and a larger initial envelope mass. Comparing the progenitors with $M_\mathrm{prog}=35\;M_\odot$ from \citet{Aguilera-Dena_2020} and \citet{Woosley_2006}, we measured $M_\mathrm{ej}=4.3\;M_\odot$ for AD035x1.0 while $M_\mathrm{ej}=10.67\;M_\odot$ in the case of  \texttt{35OC}.

Analysing the impact of the progenitor model and its rotation on the final ejecta, we found that more massive stars reach higher explosion energy because of a higher mass-infall rate that supplies a larger amount of matter to form the disc and, therefore, a higher thermal energy budget for the explosion energy. Our results also show that faster-rotating progenitors experience more energetic explosions due to a larger disc mass.

We find that the distribution of explosion energy and ejecta mass has a fair agreement with the observed distribution of SNe Ic-BL. This indicates that such an energetic explosion would be driven indeed by the disc wind in collapsars. We also find a strong correlation between the explosion energy and the binding energy of the outer layer of the star, $E_\mathrm{ol}$. We provided a function of $E_\mathrm{ol}$ to predict the explosion energy only with the information of the pre-collapse star.

As for the analysis of the $^{56}$Ni production, in our simulations, $M^\mathrm{inj}_\mathrm{ej}$ mainly determines the estimate of the total mass of $^{56}$Ni synthesized in the ejecta as it is more than one order of magnitude larger than $M^\mathrm{stellar}_\mathrm{ej, >5GK}$. Due to the predominance of $M^\mathrm{inj}_\mathrm{ej}$, for which the complete thermodynamical history is not available, we provided only a rough estimate of $M_\mathrm{ej,Ni}$. Nonetheless, the distribution of our estimate of the $^{56}$Ni mass can broadly explain the relation between $M_\mathrm{Ni}$ and $E_\mathrm{expl}$ of the observational data for stripped-envelope SNe (some of which are Ic-BL SNe) taken from \citet{Taddia2019jan} and \citet{Gomez2022dec}. We also found a correlation between the $^{56}$Ni mass and the progenitor's mass and angular velocity, i.e., more massive and faster-rotating stars produce more $^{56}$Ni. 

We also found that our results on the explosion energy and $^{56}$Ni mass agree approximately with those obtained by more detailed simulations~(\citealt{Fujibayashi:2023oyt,Dean_2024}). This suggests that our rather simple hydrodynamics model captures the essence of the explosion mechanism in the collapsar scenario and is useful for interpreting the observational data.
Yet, we found a tighter correlation of $E_\mathrm{expl}$ and $M_\mathrm{ej}$ than those of the observational data (\citealt{Taddia2019jan} and \citealt{Gomez2022dec}).
It is partially because of the limited class of progenitor structure. Another reason is likely to be the simple modeling of wind injection employed in this study (see \S~\ref{sec:Progenitor_Parameters}). We will sophisticate our current wind injection model to capture the relevant physical processes consistently and further investigate the collapsar disc wind scenario in the future.

\section*{Acknowledgements}
We want to thank David Aguilera-Dena for providing his stellar evolution model and Kengo Tomida for his help and suggestions in using \texttt{Athena++}. We are thankful for the useful and constructive discussion we had with Ayako Iahii.
This study was supported in part by Grants-in-Aid for Scientific Research of the Japan Society for the Promotion of Science (JSPS, Nos. JP22K20377 and JP23H04900).
Numerical computation was performed on Sakura cluster at Max Planck Computing and Data Facility.

\section*{Data availability}
The data will be shared on reasonable request to the corresponding author.


\bibliographystyle{mnras}
\bibliography{biblio}




\appendix

\section{Estimation of the temperature that the injected matter experiences}\label{Appendix:T_inj_matter}
In this appendix, we estimate the temperature that the injected matter experiences. As the matter is assumed to be launched from the disc, we first estimate the temperature at the typical radius of the disc $r_\mathrm{disc}$. If the internal energy is dominated by non-relativistic particles, the temperature at the radius of the disc may be estimated from the gravitational binding energy as
\begin{align}
T_\mathrm{nr} \sim \frac{G m_p M_\mathrm{BH}}{k_\mathrm{B} r_\mathrm{disc}} \sim \SI{2e11}{K} \bigg(\frac{M_\mathrm{BH}}{10M_\odot}\bigg) \bigg(\frac{r_\mathrm{disc}}{\SI{e8}{cm}}\bigg)^{-1}, \label{eq:T-nr}
\end{align}
where $m_p$ is the proton mass and $k_\mathrm{B}$ is the Boltzmann's constant. On the other hand, if relativistic particles, i.e., photons and thermally generated electron-positron pairs in high temperature, are dominant in the internal energy, the temperature is obtained by solving $aT^4/\rho \sim GM_\mathrm{BH}/r_\mathrm{disc}$ as
\begin{align}
T_\mathrm{rel} &\sim \bigg(\frac{G \rho M_\mathrm{BH}}{a r_\mathrm{disc}}\bigg)^{1/4}\notag\\
&\sim \SI{2e10}{K} \ \bigg(\frac{M_\mathrm{BH}}{10M_\odot}\bigg)^{1/4} \bigg(\frac{M_\mathrm{disc}}{0.1M_\odot}\bigg)^{1/4} \bigg(\frac{r_\mathrm{disc}}{\SI{e8}{cm}}\bigg)^{-1}, \label{eq:T-rel}
\end{align}
where $a$ is the radiation constant, and the disc density is assumed to be $\rho \sim M_\mathrm{disc}/r_\mathrm{disc}^3$. The disc temperature may be the lower value of $T_\mathrm{nr}$ and $T_\mathrm{rel}$, 
\begin{align}
T_\mathrm{disc}=\min(T_\mathrm{nr},T_\mathrm{rel}).\label{eq:Tdisc}
\end{align}
We note that $M_\mathrm{BH}$, $M_\mathrm{disc}$, and $r_\mathrm{disc}$ are functions of time, and hence, $T_\mathrm{disc}$ is also a function of time.

\begin{figure}
\centering
\includegraphics [width=0.48\textwidth]{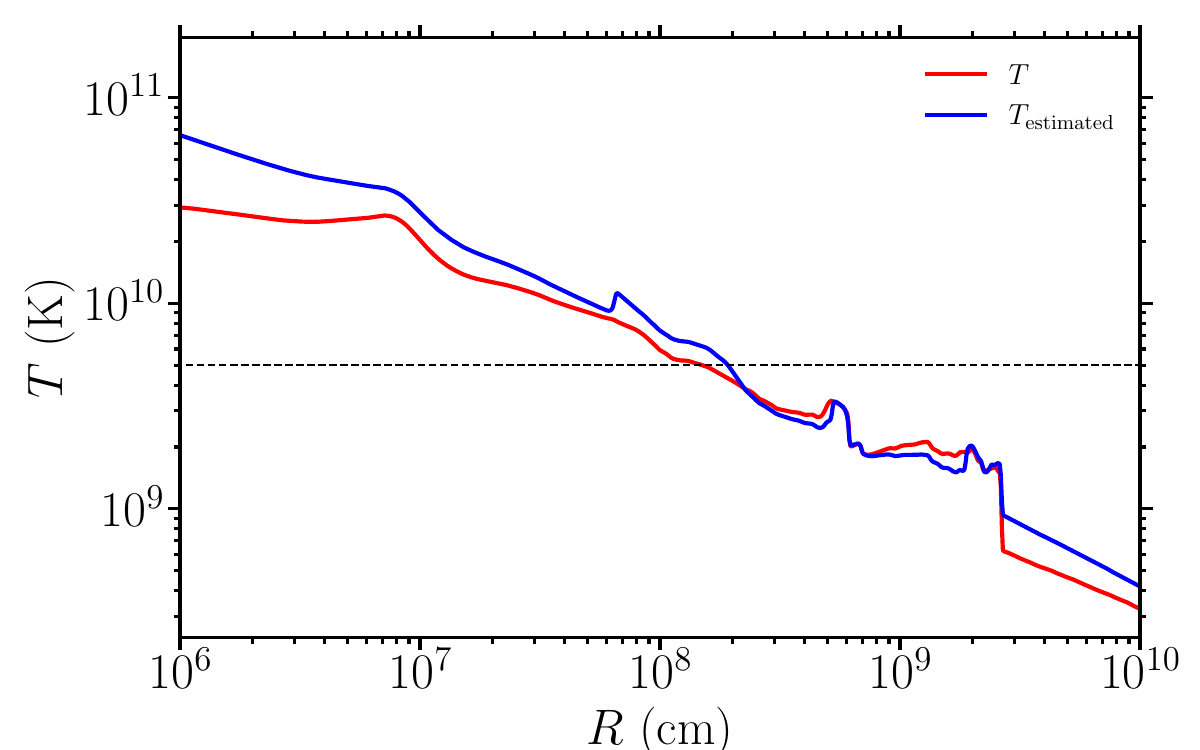}
\caption{Temperature along the equatorial direction (red line) for the model AD35-15 in \citet{Fujibayashi:2023oyt} at $t=\SI{10.3}{s}$, at which the disc outflow sets in. It is compared with the temperature estimated by Eq.~\eqref{eq:Tdisc} (blue line) with the cylindrical radius $R$, local rest-mass density $\rho$, and the BH mass $\approx 16M_\odot$ at the same time.}
\label{fig:T-estimate-AD35-15}
\end{figure}

To assess how well we estimate the disc temperature with Eq.~\eqref{eq:Tdisc}, we try to estimate the disc temperature with the same equation for a snapshot of the model AD35-15 in \cite{Fujibayashi:2023oyt}. Figure~\ref{fig:T-estimate-AD35-15} compares the temperature along the equatorial direction (red line) and the temperature estimated by Eq.~\eqref{eq:Tdisc} with the cylindrical radius and the local rest-mass density, and the BH mass (blue line) at $t=\SI{10.3}{s}$ (\SI{6}{s} after the disc formation). This corresponds to the time at which the disc wind sets in. We find that Eq.~\eqref{eq:Tdisc} estimates the actual disc temperature within a factor of two for $T\lesssim\SI{3e10}{K}$. Above this temperature, the neutrino emission can extract the internal energy in the disc evolution (viscous) timescale. Thus, the temperature is lower than the estimated value. Nevertheless, as the threshold temperature for $^{56}$Ni production is \SI{5e9}{K}, we do not have to correctly estimate the disc temperature above $\sim\SI{e10}{K}$. We thus conclude that  Eq.~\eqref{eq:Tdisc} can approximate the disc temperature with good accuracy.

\begin{figure}
\centering
\includegraphics [width=0.48\textwidth]{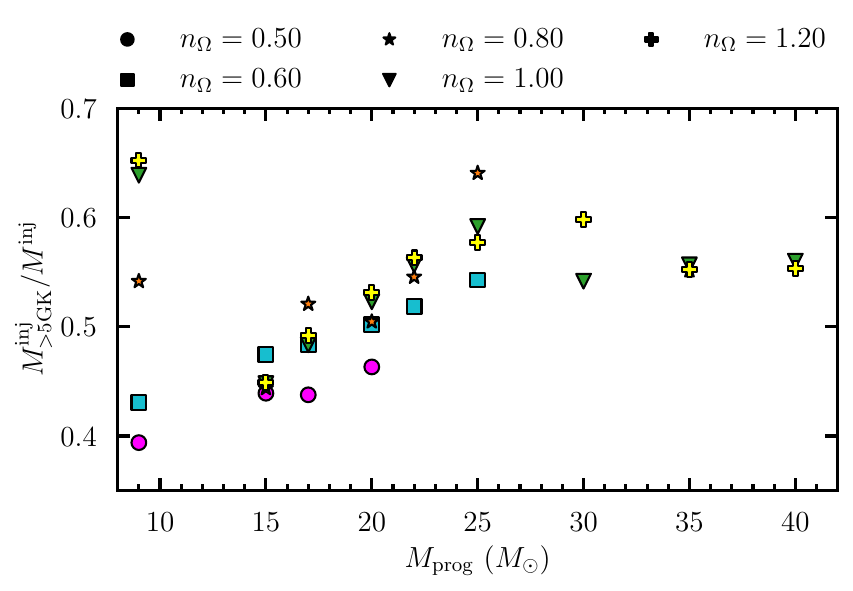}
\caption{The ratios of the $M^\mathrm{inj}_\mathrm{>5GK}$ and $M^\mathrm{inj}$ for all the models. Markers and colors distinguish the values of $n_\Omega$.
}
\label{fig:M_inj5GK_ratio}
\end{figure}

The mass of the injected matter that experiences temperature higher than \SI{5}{GK} is then calculated as
\begin{align}
M^\mathrm{inj}_\mathrm{>5GK} = \int \dot{M}_\mathrm{inj}(t) \Theta(T_\mathrm{disc}(t)-\SI{5}{GK}) dt,
\end{align}
where $\Theta$ is the Heaviside function. In Figure~\ref{fig:M_inj5GK_ratio}, the ratios of $M^\mathrm{inj}_\mathrm{>5GK}$ and $M^\mathrm{inj}$ are shown for all the models of \citet{Aguilera-Dena_2020}. We find that about half of the injected matter experiences temperature higher than \SI{5}{GK}. Thus, we may infer that a significant fraction of the injected matter is composed of $^{56}$Ni (given that its electron fraction is $\sim0.5$).

\section{Exceptional behavior of some models}\label{Appendix:exceptional_models}
\begin{figure*}
\centering
	\includegraphics [width=0.48\textwidth]{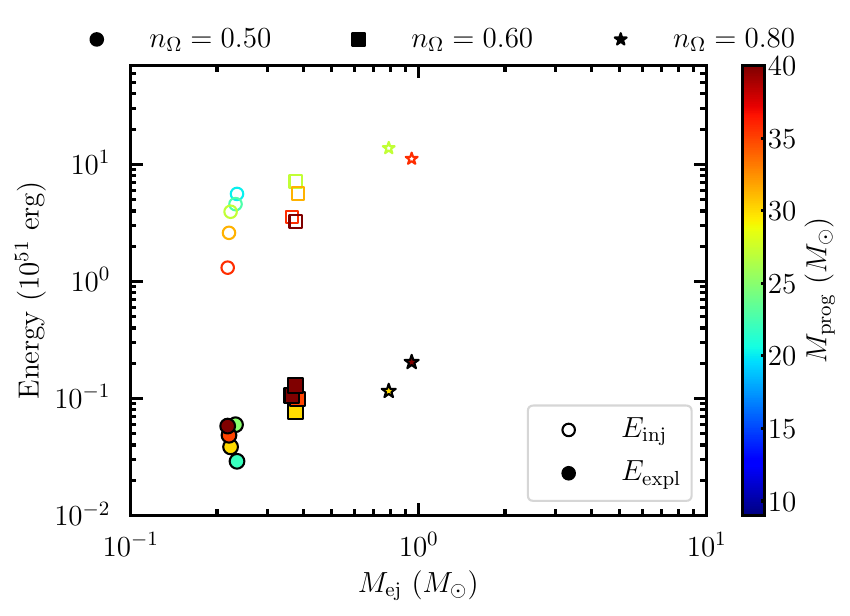}	
	\includegraphics [width=0.48\textwidth]{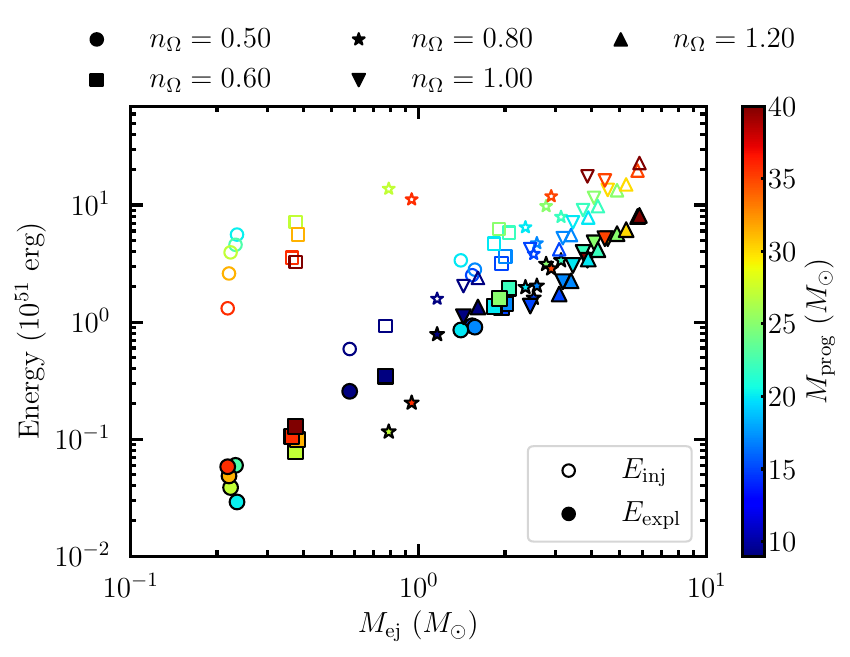}
\caption{$E_\mathrm{expl}$ (filled markers) and $E_\mathrm{inj}$ (open markers) as functions of the ejecta mass $M_\mathrm{ej}$ for all the models studied in this work (left panel) and for those showing an exceptional behavior compared to the general trend (right panel). Results for progenitors with different magnitudes of the angular velocity are distinguished by different markers, while the color indicates the progenitor mass $M_\mathrm{prog}$.}
\label{fig:Mej_Eexpl_exception}
\end{figure*}

As mentioned in \S~\ref{sec:Results}, we exclude some models from the analysis of the study because they showed exceptional behavior during the simulations. The results for these simulations are listed in Table~\ref{tab:models_exceptions}.
In the left panel of Figure~\ref{fig:Mej_Eexpl_exception}, we show the distributions of the injected and explosion energy as functions of the ejecta mass for these models only, while in the right panel, we compare them to the other progenitors. From these plots, it is evident that the values of $E_\mathrm{inj}$, $E_\mathrm{expl}$ and $M_\mathrm{ej}$ of the excluded models do not align with those of the others. They are all progenitors with $M_\mathrm{prog}>22\;M_\odot$ and they present $E_\mathrm{inj}$ higher than expected while $E_\mathrm{expl}$ and $M_\mathrm{ej}$ lower.

The reason for this behavior is that the matter with sufficient angular momentum inflows to the center and forms a centrifugally supported disc \textit{in} the computational domain ($r>r_\mathrm{in}$). As the matter that does not flow inside the inner boundary cannot be the energy source as an injected matter, the injected energy for such models is underestimated. This behavior is physically inconsistent because the viscous effect is not taken into account in the computational domain.
Models are regarded as physically consistent if the explosion sets in before the disc formation inside the computational domain. For such models, the injected matter pushes the stellar envelope and prevents it from inflowing into the vicinity of the inner boundary.

\begin{table*}
\centering
\caption{Exceptional models excluded from the analysis of the results. After the column with the model's name, from left to right, the columns list the cumulative injected energy, ejecta mass, explosion energy, average ejecta velocity, the number of tracer particles located within the ejecta, the mass of ejecta component originated from the injected matter, the ratio between the injected matter that is estimated to experience temperature higher than 5 GK and the total injected mass, the mass of ejecta component that is originated from the computational domain and experiences temperature higher than 5\,GK, along with the number of tracer particles in parenthesis, the total mass of the ejecta which experiences temperature higher than 5\,GK. 
}
\begin{tabular}{c|ccccccccc}
\hline\hline
model & $E_\mathrm{inj}$ & $M_\mathrm{ej}$ & $E_\mathrm{expl}$ & $v_\mathrm{ej}$ & $N_\mathrm{p}$ & $M_\mathrm{ej}^\mathrm{inj}$ & $M_\mathrm{>5GK}^\mathrm{inj}/M^\mathrm{inj}$ & $M^\mathrm{stellar}_\mathrm{ej,>5GK}$ ($N^\mathrm{stellar}_\mathrm{ej,>5GK}$) &  $M_\mathrm{ej,>5GK}$ \\
& ($10^{51}$\,erg) & ($M_\odot$) & ($10^{51}$\,erg) & ($10^3$\,km/s) &      & ($M_\odot$)      &       &  ($M_\odot$) &($M_\odot$)   \\  
\hline
AD022x0.5 &  5.59     &  0.22   &    0.03  &  3.53      &   120913 &  0.003 &  0.90 &  0.0230 (  9791) & 0.03\\
AD025x0.5 &  4.58     &  0.22   &    0.06  &  5.09      &   72916 &  0.001  &  0.56 &  0.00 (     1)   & 0.00056\\
AD030x0.5 &  3.94     &  0.21   &    0.04  &  4.17      &   103540 & 0.004  &  0.89 &  0.03 ( 13151)   &  0.03 \\
AD035x0.5 &  2.60     &  0.21   &    0.05  &  4.70      &   112094 &  0.002 &  0.82 &  0.05 ( 22152) & 0.05\\
AD040x0.5 &  1.31     &  0.21   &    0.06  &  5.18      &   106482 & 0.003  &  0.81 &  0.041 ( 18387)& 0.04\\
\hline

AD030x0.6 &  7.14     &  0.36   &    0.08  &  4.56      &   95468  & 0.006 & 0.91  &  0.10 ( 19466) &  0.11\\
AD035x0.6 &  5.62     &  0.38   &    0.10  &  5.11      &   139220 & 0.005 & 0.86 &  0.10 ( 29247) &  0.10\\
AD040x0.6 &  3.54     &  0.36   &    0.11  &  5.41      &   112656 & 0.002 &  0.84 &  0.10 ( 35565) & 0.10\\
\hline
AD030x0.8 &  13.76    &  0.78   &    0.12  &  3.83      &   220315 & 0.007 & 0.90  &  0.16 ( 33604) &  0.17\\
AD040x0.8 &  11.16    &  0.92   &    0.20  &  9.97      &   93655 & 0.008 &  0.89 &  0.34 ( 27650) & 0.35\\
\hline
\end{tabular}
\label{tab:models_exceptions}
\end{table*}

\section{Dependence of $E_\mathrm{expl}$ for the 35$M_\odot$ progenitor}\label{Appendix:model_35_M}
This work was motivated by our aim of better investigating the variety of explosion properties for different progenitors after the results obtained in \citetalias{LCMenegazzi}, where we instead studied different parameters of wind injection. We expected that the variation of the progenitor mass and its angular velocity would have explained the variety of the observational data. However, as discussed in \S~\ref{sec:Results:Comparison_obs}, our numerical results have a tighter correlation between $E_\mathrm{expl}$ and $M_\mathrm{ej}$ than that measured in by the observational data by  \citet{Taddia2019jan} and \citet{Gomez2022dec} (as shown in Figure~\ref{fig:Eexp_vs_Mej_w_obs_data}). A similar strong correlation between $E_\mathrm{expl}$ and $M_\mathrm{ej}$ was also found in our previous work for the model AD020x1.0 (Figure~11 in \citetalias{LCMenegazzi}); however the points of the distributions lie on lines with different slopes (see Figure~\ref{fig:Eexp_vs_Mej_w_obs_data}). This comparison suggests that the proportionality between $E_\mathrm{expl}$ and $M_\mathrm{ej}$ relates to the choice of the progenitor model and of the parameters for the wind injection. Therefore, we expect that fixing the progenitor model and studying its explosion varying the parameters of the wind injection, the explosion energy would distribute along a different trend for every progenitor star.  To confirm our speculations, we perform additional simulations using the progenitor model AD035x1.0 and varying the parameters of the wind injections sampling some values of $t_\mathrm{w}$, $t_\mathrm{acc}$, $f_\mathrm{therm}$ and $\xi^2$ among those used in \citetalias{LCMenegazzi} (see \S~\ref{sec:Method}). 
The parameters used, the results for the explosion energy, the ejecta mass, and the averaged velocity are displayed in Table~\ref{tab:models_35M}.

\begin{figure}
	\centering
	\includegraphics [width=0.48\textwidth]{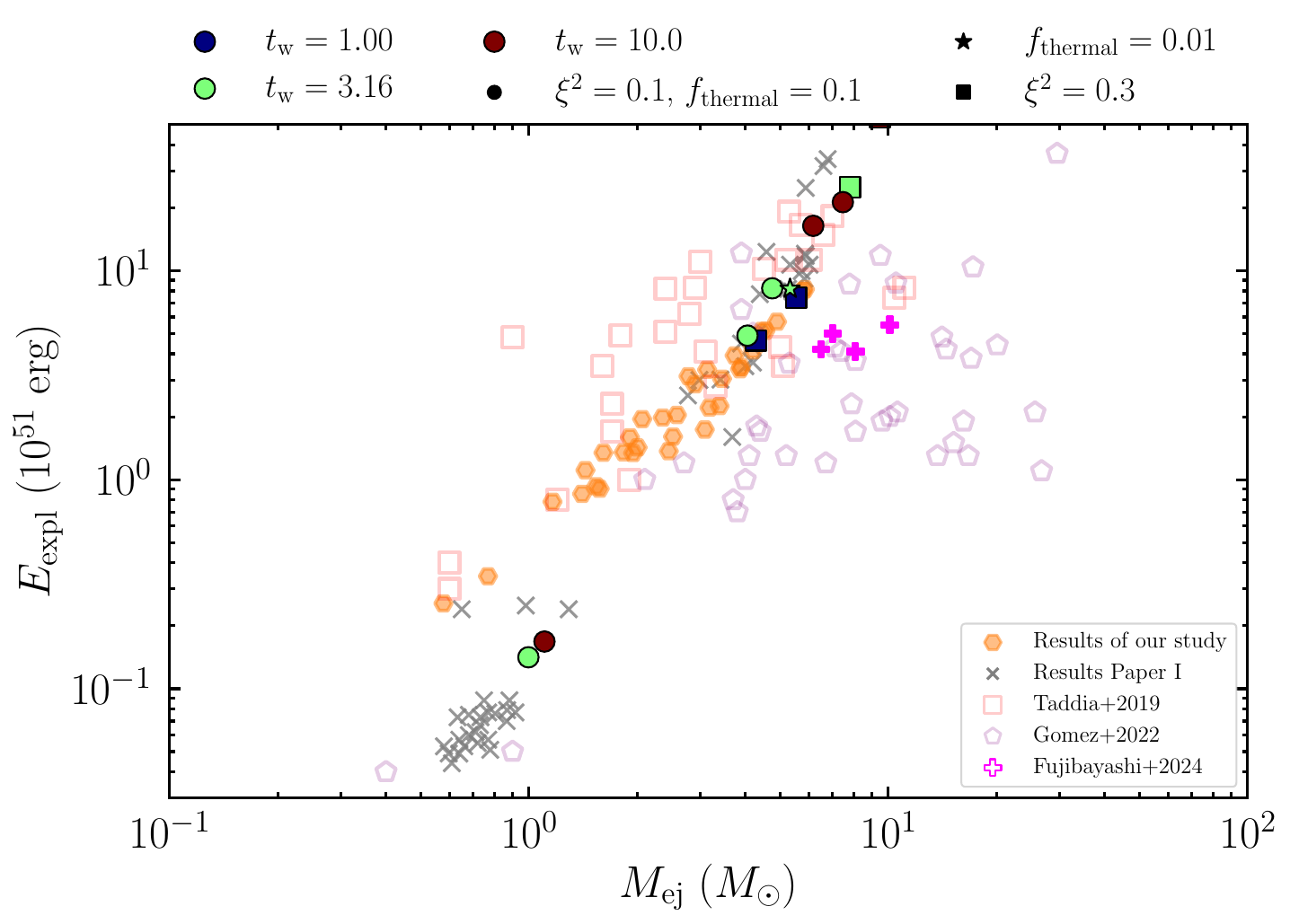}
	\caption{Wind parameter dependence of model M035x1.0 with respect to the observable pair of ejecta mass $M_\mathrm{ej}$ and explosion energy $E_\mathrm{expl}$. The color distinguishes the wind time scale $t_\mathrm{w}$. The orange hexagonal markers show the result of the study presented in this paper, while the gray x-markers represent the results obtained in \citetalias{LCMenegazzi} for the model AD020x1.0. The open markers display the observational data for stripped-envelope SNe, some of which are Type Ic-BL SNe, taken from \citet{Taddia2019jan} and \citet{Gomez2022dec}. The magenta plus-sign denotes the results obtained in a general relativistic neutrino-radiation viscous-hydrodynamics simulation with the same progenitor AD035x1.0 by \citet{Fujibayashi:2023oyt}.
	}
    \label{fig:Eexp_vs_Mej_M35}
\end{figure}

In Figure~\ref{fig:Eexp_vs_Mej_M35}, we show in the $E_\mathrm{expl}$-$M_\mathrm{ej}$ plane the distribution of the models with different parameters of the wind injection for the progenitor AD035x1.0 (filled markers). We also show the results obtained for the study presented in this work (see \S~\ref{sec:Results} for different progenitor models with the parameters of the wind injection fixed (orange hexagonal markers) and the outcomes obtained in \citetalias{LCMenegazzi} with gray $\times$-markers. We additionally display the observational data from \citet{Taddia2019jan} and \citet{Gomez2022dec} (open markers) and the results obtained by \citet{Fujibayashi:2023oyt} for the same progenitor AD035x1.0.

Comparing the distribution of the explosion energy of AD035x1.0  with that of AD020x1.0, they show two slightly different trends, i.e., they lie on lines with different slopes. This supports our speculation that by varying the progenitor structure and sophisticating the wind injection model, we may be able to reproduce a wider variety of the observational data. However it is still impossible to reproduce the observational data with $M_\mathrm{ej} \gtrsim 10M_\odot$ because our choice of the progenitor models are not appropriate for this purpose.

\begin{table*}
\centering
\caption{Model description and key results for model M035x1.0 studied with different parameters for the wind injection. From left to right, the columns contain the wind time scale, the ratio of the accretion and wind time scales, the squared ratio of the asymptotic velocity of injected matter to escape velocity of the disc, the internal to kinetic energy ratio of injected matter, cumulative injected energy, ejecta mass, explosion energy, and average ejecta velocity.
}
\begin{tabular}{l|cccccccc}
\hline\hline
model &$t_\mathrm{w}$ & $t_\mathrm{acc}/t_\mathrm{w}$ & $\xi^2$ & $f_\mathrm{therm}$ & $E_\mathrm{inj}$ & $M_\mathrm{ej}$ & $E_\mathrm{expl}$ & $v_\mathrm{ej}$ \\ 
&(s) & & & & ($10^{51}$\,erg) & ($M_\odot$) & ($10^{51}$\,erg) & ($10^3$\,km/s)  \\ 
\hline
M35\_1\_3.16\_0.1\_0.10 &      1 &     3.16 & 0.1 & 0.10 &   15.25    &  4.29  &  4.64  &    10.43\\
M35\_1\_10\_0.1\_0.10 &        1 &     10   & 0.1 & 0.10 &   18.06    &  5.53  &  7.45  & 11.63  \\
M35\_3.16\_1\_0.1\_0.10 &      3.16 &  1 & 0.1 & 0.10 &      0.52     &  0.99  &  0.14  & 3.78   \\
M35\_3.16\_3.16\_0.1\_0.10 &   3.16 &  3.16 & 0.1 & 0.10 &   16.29     & 4.06  &  4.89  & 11.01   \\
M35\_3.16\_10\_0.1\_0.10 &     3.16 &  10   & 0.1 & 0.10 &   21.13    &  4.75  &  8.23  &  13.20  \\
M35\_3.16\_10\_0.3\_0.10 &     3.16 &  10   & 0.3 & 0.10 &   40.20    &  7.82  &  25.02 &  17.94\\
M35\_3.16\_10\_0.1\_0.01 &     3.16 &  10   & 0.1 & 0.01 &   20.53    &  5.33  &  8.19  &  12.44   \\
M35\_10\_1\_0.1\_0.10 &        10 &    1    & 0.1 & 0.10 &   1.89     &  1.10  &  0.17  &   3.91 \\
M35\_10\_3.16\_0.1\_0.10 &     10 &    3.16 & 0.1 & 0.10 &   28.88    &  6.18  &  16.39  &  16.33  \\
M35\_10\_3.16\_0.3\_0.10 &     10 &    3.16 & 0.3 & 0.10 &   69.53    &  9.48  &  54.25  &  23.99  \\
M35\_10\_10\_0.1\_0.10 &       10 &    10 & 0.1 & 0.10 &     35.15    &  7.47  &  21.27  &  16.92  \\
M35\_10\_10\_0.3\_0.10 &       10 &    10 & 0.3 & 0.10 &     88.43    &  9.52  &  72.27  &  27.62  \\
\hline
\end{tabular}
\label{tab:models_35M}
\end{table*}

\label{lastpage}
\end{document}